%% file: paper.tex
\documentclass[acmtog, authorversion]{acmart}
\input{sec/0_preamble.tex}
\input{sec/tweaks.tex}

\input{sec/math.tex}

\begin{document} 
\input{sec/0_front}
\input{fig/teaser/item}
\input{sec/0_abstract}
\maketitle
\thispagestyle{empty} %
\input{sec/1_intro.tex}
\input{sec/2_related}

\input{sec/3_rods} %
\input{sec/4_solver} %
\input{sec/5_rodsolver} %
\input{sec/6_optimization} %
\input{sec/7_engine} 
\input{sec/8_muscles}
\input{sec/10_conclusion}
\bibliographystyle{ACM-Reference-Format}
\bibliography{paper}
\clearpage
\appendix

\input{sec/12_appendix_energies}

\input{sec/11_appendix}
\input{sec/13_appendix_eval}
\clearpage
\end{document}

%% file: sec/0_preamble.tex
\setcitestyle{square}

\usepackage[ruled]{algorithm2e} %

\SetAlFnt{\small}
\SetAlCapFnt{\small}
\SetAlCapNameFnt{\small}
\SetAlCapHSkip{0pt}
\IncMargin{-\parindent}

\renewcommand\footnotetextcopyrightpermission[1]{}
\setcopyright{none}

%% file: sec/tweaks.tex
\usepackage{overpic}
\usepackage{amssymb}
\usepackage{amsmath}
\usepackage{graphicx}
\usepackage{wrapfig}
\usepackage{float} %
\usepackage{caption} %
\usepackage{microtype}
\usepackage{booktabs} %
\usepackage{dirtytalk}
\usepackage{mathtools}
\usepackage{cases}

\usepackage{color}
\definecolor{turquoise}{cmyk}{0.65,0,0.1,0.3}
\definecolor{purple}{rgb}{0.65,0,0.65}
\definecolor{dark_green}{rgb}{0, 0.5, 0}
\definecolor{orange}{rgb}{0.8, 0.6, 0.2}
\definecolor{red}{rgb}{0.8, 0.2, 0.2}
\definecolor{darkred}{rgb}{0.6, 0.1, 0.05}
\definecolor{blueish}{rgb}{0.0, 0.3, .6}
\definecolor{light_gray}{rgb}{0.7, 0.7, .7}
\definecolor{pink}{rgb}{1, 0, 1}
\definecolor{greyblue}{rgb}{0.25, 0.25, 1}

\newcommand{\CIRCLE}[1]{\raisebox{.5pt}{\footnotesize \textcircled{\raisebox{-.6pt}{#1}}}}

\newcommand{\hidden}[1]{{}} %

\newcommand{\todo}[1]{{\color{red}#1}}

\newcommand{\ADDRESSED}[1]{{}}

\newcommand{\at}[1]{{\color{blueish}#1}}

\newcommand{\Figure}[1]{Figure~\ref{fig:#1}}
\newcommand{\Eq}[1]{Equation~\ref{eq:#1}}
\newcommand{\eq}[1]{(\ref{eq:#1})}
\newcommand{\Equation}[1]{Equation~\ref{eq:#1}}
\newcommand{\Sec}[1]{Sec.~\ref{sec:#1}}
\newcommand{\Section}[1]{Section~\ref{sec:#1}}
\newcommand{\Appendix}[1]{Appendix~\ref{app:#1}}

\providecommand{\citeN}[1]{\todo{Blah et al.~\cite{#1}}}
\providecommand{\citet}[1]{\todo{Blah et al.~\cite{#1}}}

\newcommand{\brief}[1]{} %
\newcommand{\ignore}[1]{} 
\usepackage{currfile}

\usepackage{comment}
\specialcomment{DRAFT}{\color{red}}{\color{black}}

\usepackage{xcolor}

\renewcommand{\paragraph}[1]{{\textbf{#1.}}}
\usepackage{pdfcomment}
\providecommand{\AndreaComment}[1]{\pdfmargincomment[color=red,icon=Note]{#1}}
\providecommand{\AndreaComment}[1]{}

\usepackage{amsthm}
\newtheorem{defn}{Definition}[section]

\usepackage{pifont}

%% file: sec/math.tex
\usepackage{amsmath}

\renewcommand{\vec}[1]{\mathbf{#1}}
\newcommand{\mat}[1]{\mathbf{#1}}

\newcommand{\onehalf}{.5}

\newcommand{\pill}{pill}

\newcommand{\intD}{\int_{D}}

\newcommand{\atrest}[1]{\bar{#1}}

\newcommand{\rot}{\mat{R}}

\newcommand{\stiff}{\mat{K}}
\newcommand{\pot}{\mat{W}}
\newcommand{\darboux}{\boldsymbol{\Omega}}
\newcommand{\ct}{\vec{c}}
\newcommand{\rest}{\bar{\vec{p}}}
\newcommand{\deform}{\vec{p}}
\newcommand{\predict}{\hat{\deform}}
\newcommand{\q}{\vec{q}}
\newcommand{\p}{\vec{p}}
\newcommand{\f}{\vec{f}}
\newcommand{\x}{\vec{x}}
\newcommand{\dx}{\Delta \x}

\newcommand{\rotu}{\vec{u}}
\newcommand{\rotv}{\vec{v}}
\newcommand{\rotw}{\vec{w}}

\DeclareMathOperator*{\argmin}{arg\,min}

\DeclareMathOperator*{\matsum}{sum}
\newcommand{\gradient}{\nabla}

%% file: sec/0_front.tex
\title{VIPER: Volume Invariant Position-based Elastic Rods}
\author{Baptiste Angles}
\affiliation{Google}
\affiliation{Universit\'e de Toulouse}
\affiliation{Electronic Arts / SEED}
\affiliation{University of Victoria}
\author{Daniel Rebain}
\affiliation{Google}
\affiliation{University of Victoria}
\author{Miles Macklin}
\affiliation{NVIDIA}
\affiliation{University of Copenhagen}
\author{Brian Wyvill}
\affiliation{University of Victoria}
\author{Loic Barthe}
\affiliation{Universit\'e de Toulouse and IRIT / CNRS}
\author{JP Lewis}
\author{Javier von der Pahlen}
\affiliation{Electronic Arts / SEED}
\author{Shahram Izadi}
\author{Julien Valentin}
\author{Sofien Bouaziz}
\affiliation{Google}
\author{Andrea Tagliasacchi}
\affiliation{Google Research}
\affiliation{University of Waterloo}
\affiliation{University of Victoria}
\authorsaddresses{}
\renewcommand{\shortauthors}{Angles et al.}


\ccsdesc[500]{Computing methodologies~Physical simulation}
\ccsdesc[500]{Computing methodologies~Volumetric models}
\ccsdesc[100]{Computing methodologies~Collision detection}

\keywords{Cosserat rods, soft-body deformation.}

%% file: fig/teaser/item.tex
\begin{teaserfigure}
\begin{minipage}[t]{\columnwidth}
\begin{overpic} 
[width=\columnwidth]
{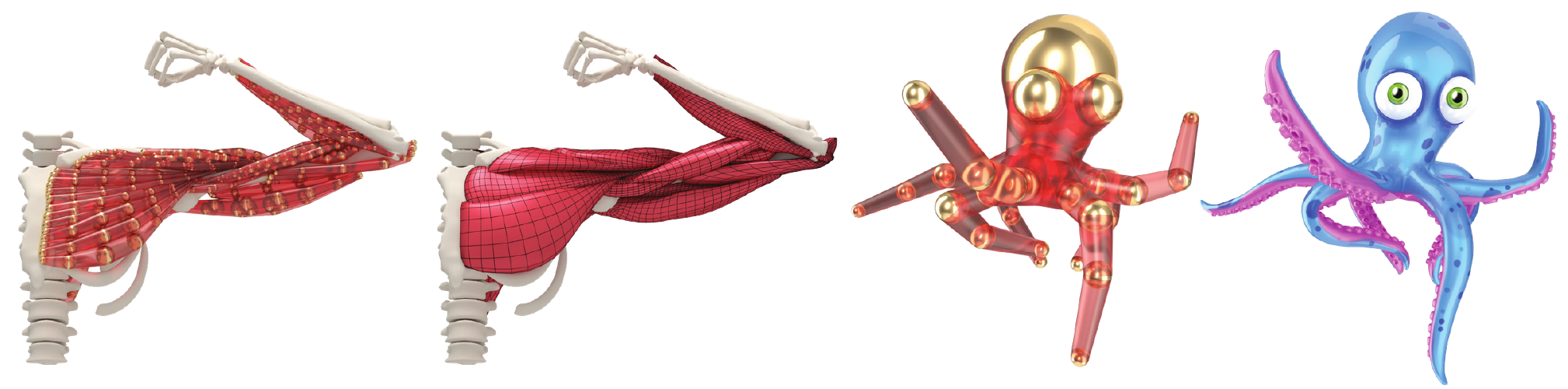}
\put(12, 1){\small{(a)}}
\put(39, 1){\small{(b)}}
\put(63, 1){\small{(c)}}
\put(85.5, 1){\small{(d)}}
\end{overpic}
\caption{
(a) We compactly model muscles as a collection of generalized rods, where volume conservation is expressed by a radius function defined on curve's vertices -- vis sphere's radii.
(b) The rods create a subspace on which physics is solved, and its effects later propagated to the muscle mesh via linear blend skinning;
please see the animation in our \textbf{supplemental video}.
The anatomical model is courtesy of Ziva Dynamics.
(c) We show how rods and/or bundles can be skinned to a surface mesh to drive its deformation (d), resulting in an alternative to cages for real-time volumetric deformation.
}
\label{fig:teaser}
\end{minipage}
\end{teaserfigure}
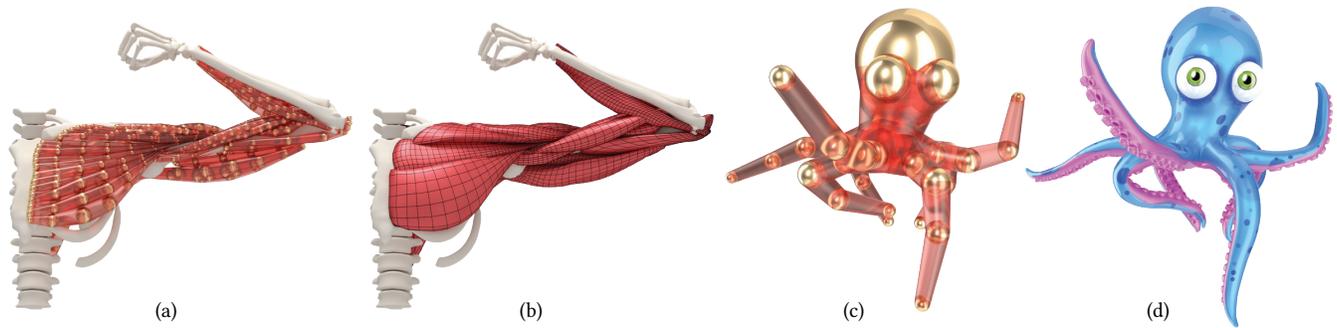

%% file: sec/0_abstract.tex
\begin{abstract}
We extend the formulation of position-based rods to include elastic \emph{volumetric} deformations. We achieve this by introducing an additional degree of freedom per vertex -- isotropic scale (and its velocity).
Including scale enriches the space of possible deformations, allowing the simulation of volumetric effects, such as a reduction in cross-sectional area when a rod is stretched.
We rigorously derive the continuous formulation of its elastic energy potentials, and hence its associated position-based dynamics (PBD) updates to realize this model, enabling the simulation of up to 26000~DOFs at 140~Hz in our GPU implementation.
We further show how rods can provide a compact alternative to tetrahedral meshes for the representation of complex muscle deformations, as well as providing a convenient representation for collision detection.
This is achieved by modeling a muscle as a \emph{bundle} of rods, for which we also introduce %
a technique to automatically convert a muscle surface mesh into a rods-bundle.
Finally, we show how rods and/or bundles can be skinned to a surface mesh to drive its deformation, resulting in an alternative to cages for real-time volumetric deformation.
The source code of our physics engine will be openly available\footnote{\href{https://vcg-uvic.github.io/viper/}{https://vcg-uvic.github.io/viper/}}.
\end{abstract}

%% file: sec/1_intro.tex
\input{fig/fibers+ziva/item.tex}
\section{Introduction} 
In recent years, the computer graphics community has invested exceptional efforts in adapting the (non real-time) physical simulation algorithms at the core of cinematic special effects (e.g.:~\cite{ziva}) to the realm of (real-time) interactive applications (e.g.: games, AR/VR). Many of these advancements have been possible thanks to a new class of physics solver, pioneered by \citeN{pbd}, realized on top of \emph{Verlet}-class integrators \cite{pbdstar}.
These \emph{position-based} solvers are capable of elegantly modeling constrained Newtonian dynamics, including rigid-body, cloth, ropes, rods and fluids in a unified framework. A~primary example is the NVIDIA FLEX system \cite{flex}, capable of modeling complex and varied physical phenomena in real-time by leveraging modern GPU hardware.

\paragraph{Rods with volume}
Within this technological landscape, of particular relevance to our work is the modeling of elastic ``rods''~\cite{strands,rods,pbdrods,pbdcorods}.
These models extend ``ropes'' by augmenting each segment composing the rod with an orthogonal coordinate frame, hence allowing the modeling of \emph{torsion} on top of stretching/bending.
Our VIPER rods extend these formulations by accounting for volume preservation, a phenomena not modeled by existing position-based rod models. 
Many interesting phenomena require this constraint (e.g. soft-bodies, fluids).
For example, water is the largest constituent of most animal tissues ($\approx\!80\%$ in muscles) hence modeling quasi-incompressible phenomena is of critical importance to achieve believable motion.
We address this problem by adding a \emph{per-vertex scaling} degree of freedom -- a measure of the local rod cross-section -- and optimizing for this quantity within the physics solve.
Our rod segments are \emph{hybrid} surface representations, they are \emph{explicitly} parameterized by the position and scale/radius of their vertices, but their surface is defined \emph{implicitly}.
This hybrid structure makes them particularly well suited for efficient collision detection/resolution \cite{green2010particle}.

\paragraph{Anatomical modeling}
Such a physical model not only satisfies our fixation in efficiently simulating rubber bands, but has immediate applications towards the modeling of muscles.
As illustrated in \Figure{fibers}, \emph{striated skeletal muscles}\footnote{From now on, we will refer to ``striated skeletal muscle'' simply as ``muscles''.} in human bodies can be represented as a collection of fibers surrounded by connective tissue (i.e. fasciae). 
Simulating these muscle types efficiently is a fundamental problem, as they represent from $36\%$ to $42\%$ of the average human body mass.
In this paper, we propose to efficiently model muscles as a structured collection of volume-preserving rods. 
This new model can also be interpreted as a generalization of the static volumetric primitives in \emph{Implicit Skinning}~\cite{impski}, where skin can then be efficiently modeled as a triangular mesh sliding on an implicit iso-surface defined by our fibers.
The VIPER primitive is designed to be integrated with other existing components to produce a complete character representation. These include representations of the skeleton, fat, skin,  etc.

\paragraph{Volumetric simulation}
In the industry, volume-preserving simulation is typically performed by discretizing the interior of the object with tetrahedra or using an approximating cage.
However, to the best of our knowledge, even modestly sized models require a lengthy preprocessing (e.g. a large scale eigen-decomposition for computing the deformation modes; see \cite{Barbic:2005}) before real-time simulation becomes possible.
For example, while visually striking, computing the simulation in \Figure{ziva} requires a two-pass optimization (\CIRCLE{1} muscle, \CIRCLE{2} skin). 
\CIRCLE{1} Muscles are discretized with $51k$ tets sharing $21k$ vertices, and each of their $4$ steps of simulation requires $3.2$ seconds.  
\CIRCLE{2} Skin ($78k$ vertices) is simulated as cloth layered on top of fat (having $68k$ tets sharing $18k$ vertices), where each of the necessary $4$ substeps of simulation requires $\approx 35$ seconds of compute. Overall, this cumulates to $\approx 2.5$ minutes of compute/frame.
Offline simulation can be exploited to learn sub-spaces, which then enables dynamic deformations in real-time; see \cite{pssd}.
However, once learnt, the dynamic behavior of the model is ``baked''. Clearly, this is an obstacle towards the ultimate goal of truly interactive physics simulation, and, consequently, interactive modeling.
The VIPER primitive not only allows the real-time volumetric simulation of complex anatomical structures, but also provides a viable alternative to cages as a compact control structure for soft-body deformations.

\paragraph{Automatic fiber-bundle modeling}
Rod-based representations of muscle fascia are not commonly available -- typical asset databases contain tetrahedral mesh models instead.
While artists sometimes model main characters to the level of interior muscles, this effort is expensive and not justified
for background characters. Thus we also introduce a technique to convert existing assets with minimal user intervention.
Our solution builds fiber bundles by first creating a set of slices through the muscle then performing an iterative optimization to interpolate these slices with a given number of rods such that they approximate the input surface well.

\paragraph{Contributions}
Our fundamental contribution is the design of a novel real-time physics engine for soft-body dynamics. Our system presents several sub-contributions:
\vspace{-\parskip}
\begin{itemize}
\setlength\itemsep{0em}
\item We enrich position-based solvers by introducing a new volume-preserving cosserat rods model and associated constraints.
\item We demonstrate how these primitives, when assembled into fiber-bundles, are effective in efficiently modeling muscles.
\item We propose a technique for conveniently creating fiber-bundles models from existing simulation assets.
\item We introduce the use of VIPER rods as efficient deformation proxies for soft-body deformation.
\end{itemize}

%% file: fig/fibers+ziva/item.tex
\begin{figure*}
\begin{minipage}[t]{\columnwidth}
\begin{overpic} 
[width=\columnwidth]
{\currfiledir/litem.pdf}
\end{overpic}
\caption{
(top)
Skeletal striated muscle as a collections of nested fascicles, fibres, and myofibrils; base image courtesy of~\protect\cite{lee_tog10}.
(bottom)
We abstract the fascicles as a collection of rods.
These can be overlapping, and their rest-pose structure is controlled by \emph{shape-matching} constraints.
}
\label{fig:fibers}
\end{minipage}%
$\hfill$
\begin{minipage}[t]{\columnwidth}
\begin{overpic} 
[width=\columnwidth]
{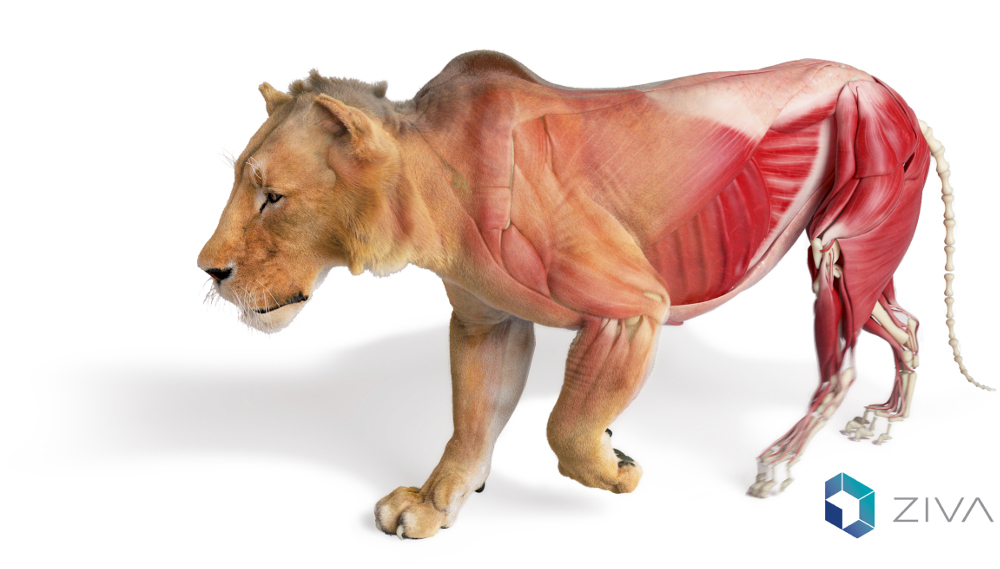}
\end{overpic}
\caption{
Bio-mechanically accurate simulation of volumetric anatomical structure is the most effective way to simulate secondary motions (e.g. skin sliding on muscles) and deliver true realistic appearance to dynamic virtual characters; image courtesy of \cite{ziva}.
}
\label{fig:ziva}
\end{minipage}
\end{figure*}

%% file: sec/2_related.tex
\section{Related Work}
\label{sec:related}
We overview the literature from different angles.
We recap example-based modeling frameworks that are commonly used in digital production, as well as recent efforts towards the use of simulated anthropomorphic models. 
We also review methods that attempt to~``emulate'' them via geometric processes, and finally processes to calibrate a given model to a target.
\paragraph{Example-based deformation}
Digital characters are often modeled via their skin (i.e. \emph{skinning}), with no consideration of underlying volumetric structures, often resulting in non-physically realistic effects such as the \emph{candy wrapper} problem of (LBS) Linear Blend 
Skinning \cite[Fig.3]{skinning}.
While these artifacts can be resolved~\cite{dqs,binhleCoR}, skinning solutions lack details such as tendons, muscle bulges, wrinkles, and volume preservation.
Example-based approaches such as (PSD) \emph{Pose-Space Deformation}~\cite{psd,kurihara04} and \emph{BlendShapes}~\cite{blendshapes} interpolate 
artist-sculpted shapes to \emph{emulate} all these effects.
However, \say{\emph{producing effects such as skin sliding over underlying structures, or the collision effects visible parts of the body press together, can require considerable skill and weeks to months of sculpting depending on the required quality}} \cite{artistquote}.
Such data-driven models can be learnt from measurements for both static~\cite{smpl}, as well as dynamic~\cite{dyna} humans, but they hardly generalize outside of their corresponding training domains.

\input{fig/cosserat/item.tex}

\paragraph{Physically-based anthropomorphic models}
Physically-based simulation of characters has a long history \cite{terzopoulos90,scheepers97,sifakis05} but, due to its high computational cost, it has only recently began to see practical use.
For skeletal muscle deformation, \citeN{lee_tog10} provides an excellent overview of the field, in regards to which~\citeN{saito_sig15}, with its ability to reach \emph{near-interactive runtime}, can be considered the state-of-the-art. 
Similarly to blendshape generation, training data can be exploited to generate efficient low-dimensional simulations~\cite{pssd,examplemat,projdyn,ichim2017phace}, enabling physically-based digital characters in production settings~\cite{avatar,ziva}.
A shortcoming of these methods is the requirement that the training set encompasses samples of all configurations of the object/character that will be needed.
Physically based approaches also permit a decomposition into \emph{layers} -- skin, muscle, fat, and bones -- enabling appropriate algorithms to be used for each.
Highly relevant to our work is the simulation of skin layered over volumetric primitives pioneered by~\citeN{thinskin}, and its realization in commercial software~\cite{vital}, as well as other existing research \cite{saitoskinslide}, industry \cite{ziva}, and proprietary solutions~\cite{avatar} to this complementary problem.
Recently, \citeN{romeo_ceig18} proposed the use of PBD for muscle simulation, but its $\approx\!40$s/frame of processing time makes it unsuitable to interactive applications.

\paragraph{Non physically-based anthropomorphic models}
Recent efforts have been made to create alternative representations for sub-skin volumetric models (i.e.  representations of muscle, fat, and bones).
For example, Maya~Muscle~\cite{mayamuscle} represents muscles via NURBS that drive the skin via LBS, whose volume is artist-driven in a PSD fashion.
However, due to the \emph{explicit} nature of NURBS, collisions are expensive, hence the performance of the system does not scale well in the complexity of the model.
Rather than driving skin via LBS, \emph{Implicit~Skinning}~\cite{impski} lets it slide on top of implicit surfaces via optimization.
These surfaces are defined by blending components that abstract entire body parts~(i.e. union of bone, muscle, and fat).
In contrast to our work, note that this model is purely kinematic~(i.e. no physics).
Another relevant class of methods simplifies anthropomorphic components even further.
For example representing an entire arm as two \emph{tapered capsules} is advantageous for arm~vs.~cloth collision detection~\cite[Cloth/Collision]{physx}.
Sphere-Meshes generalize these representations, and have recently been used to approximate geometry~\cite{sphmsh}, and track its movement in real-time~\cite{hmodel}.

\paragraph{Calibrating anthropomorphic (volumetric) models}
\citeN{ali_tog13} pioneered the transfer of anatomical structures from a template to a target human via approximate deformation models of soft tissues, and \citeN{zhu_eg15} calibrated these models from a set of RGBD images capturing a human in motion. 
Analogously to these methods, physically inspired models~\cite{saito_sig15} can be calibrated to a set of 3D surface scans~\cite{kadlecek_siga16}.
Of particular relevance to our method is the fiber estimation technique pioneered by~\citeN{choi_plos13} employed in~\cite[Sec.~3.1.1]{saito_sig15}.
While \citeN{saito_sig15} is interested in deriving the anisotropic deformation frame, we require an explicit decomposition of the muscle in fiber bundles.

\paragraph{Cosserat Rods}
Since being introduced to computer graphics by \citeN{strands}, Cosserat rods have been the subject of many efforts to achieve efficient and accurate simulation of thin materials using a variety of discretizations and energy formulations \cite{sueda2011large,rods,exact_2011,bergou2008discrete, bertails2006super, soler2018cosserat}.
\citeN{bergou2010discrete} model volume conservation in rods by updating the radius of segments based on changes in length.
In pursuit of real-time performance, rods have been simulated within a PBD framework \cite{pbdrods,pbdcorods}, a capability which we extend by further modeling volume preservation.
Finally, note rods have been applied to muscle-skeletal simulation~\cite{sueda2008musculotendon}, although to represent tendons and without consideration of volumetric effects.

%% file: fig/cosserat/item.tex
\begin{figure*}[t]
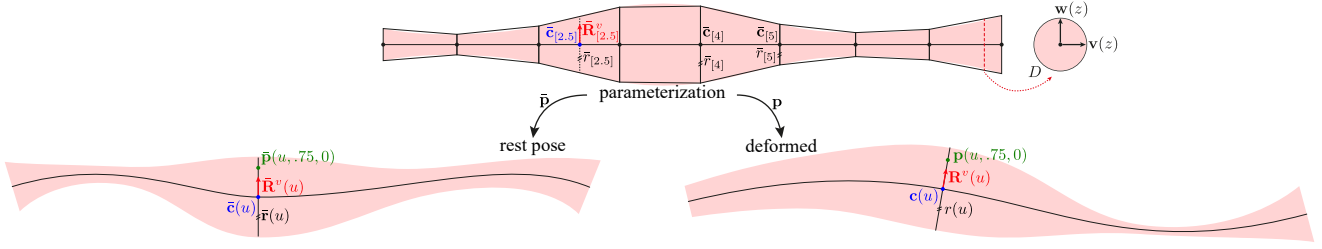

\centering
\begin{overpic} 
[width=\linewidth]
{\currfiledir/fig.pdf}
\end{overpic}
\caption{
(top) The parameterization of a VIPER rod, and its discretization. (left) Its rest configuration, and (right) a deformed configuration.
}
\label{fig:cosserat}
\end{figure*}

%% file: sec/3_rods.tex
\section{Generalized Rod Parameterization}
\label{sec:rods}
Our physical model of Cosserat rods consists of a smooth parametric curve in 3D space $\vec{c}(z)\!:\![z_0, z_1] \to \mathbb{R}^3$.
An orthogonal frame $\mat{D}(z) \in \mathbb{R}^{3 \times 3}$ is attached to every point $\ct(z) \in \mathbb{R}^{3}$ on the curve.
The orthogonal frame $\mat{D}(z) = s(z)\rot(z)$ is a combination of a uniform scale $s(u)$, and an orthonormal matrix $\rot(z)=[\rotu(z),\rotv(z),\rotw(z)]$; see \Figure{cosserat}.
Note that our model generalizes classical elastic rods~\cite{rods,pbdcorods}, as in those models the scale is kept \emph{constant} along the curve.
As shown in \Figure{cosserat}, any point in the parametrized volume of the rod can be transformed to the rest configuration by a function $\rest(x, y, z)\!:\!\mathbb{R}^3 \to \mathbb{R}^3$:
\begin{equation}
\label{eqn:rest_parametrization}
\rest(x,y,z) \equiv \rest(\q, z)=\atrest{\ct}(z) + \bar{s}(z)\atrest{\rot}(z) \q
\end{equation}
where $\q = [x, y, 0]^T$ is a point on a disc $D(z)$ of radius $r(z)$, aligned with the $xy$ plane, and centered at its center of mass.
Similarly, a second function maps the rod from parameterization to its deformed configuration:
\begin{equation}
\label{eqn:deformed_parametrization}
\deform(x,y,z) \equiv \deform(\q, z)=\ct(z) + s(z)\rot(z) \q
\end{equation}

%% file: sec/4_solver.tex
\section{Variational Implicit Euler Solver}
\label{sec:variational_implicit_euler}
Our solver is based on the variational form of implicit Euler integration~\cite{martin_tog11}. The physical model evolves through a discrete set of time samples, with simulation step size $h$. At time $t$ the deformed position is defined as $\deform_t(x, y, z)$ and the velocity as $\dot{\deform}_t(x,y,z)$. The rest pose is defined as $\rest(x,y,z) = \deform_0(x,y,z)$. The mass $m$ is assumed to be uniform over the rod. 
The sum of the external forces is denoted as $\f_{\text{ext}}(x,y,z)$. We will now drop the indexing $(x,y,z)$ whenever possible to improve readability.
We consider position dependent internal forces such that the sum of the internal forces is 
\begin{equation}
\vec{f}_{\text{int}}(x,y,z) = - \tfrac{1}{2}{\textstyle \sum}_i \nabla_{\vec{p}} \| \pot_i(\deform, \rest) \|^2_{\stiff_i},
\end{equation}
where $\pot_i(\deform, \rest)$ is a potential energy function, and  $\stiff_i$ is a matrix of stiffness parameters which we assume to be uniform over the rod, and the notation $\|\vec{x}\|_{\mat{A}}^2$ means $\vec{x}^T\mat{A}\vec{x}$.
We can then write implicit Euler as an optimization describing the compromise between an inertia potential and the elastic potentials:
\begin{equation}
\min_{\{\ct_t, s_t, \rot_t\}} \int_{z_0}^{z_1} \!\!\! \iint\limits_{D(z)} \!\!\!
\underbrace{\tfrac{m}{2h^2}\|\deform_{t} - \predict_t\|_2^2}_{inertia} 
+ 
\underbrace{\tfrac{1}{2}{\textstyle \sum}_i\| \pot_i(\deform_t, \rest) \|_{\stiff_i}^2}_{elastic} dx\,dy\,dz
\label{eq:variational_implicit}
\end{equation}
With $\predict_t$ we indicate the \emph{inertial prediction} for $\deform_t$, i.e., its next position in absence of internal forces:
\begin{align}
\predict_t  =  \deform_{t-1} + h\dot{\deform}_{t-1}+ \tfrac{h^2}{m}\vec{f}_{\text{ext}},
\end{align}
where $\dot{\deform}(\q, z) = \dot{\ct}(z) + \left( \dot{s}(z)\rot(z) + s(z)\dot{\rot}(z) \right) \q$.

\input{fig/static+dynamic/item}

\paragraph{Discretization}  We discretize the curve in the parametrized domain using a set of $m + 1$ points $\{z_{[0]}, \hdots, z_{[m]}\}$ connected using $m$ \emph{piecewise linear elements} of length $\{l_{[1]}, \hdots, l_{[m]}\}$; see~\Figure{cosserat}.
We can approximate the curve integral by integrating over these piecewise linear elements using the midpoint rule.
For the integration we also define a set of $m$ midpoints $\{z_{[{\onehalf{}}]}, \hdots, z_{[{m - \onehalf{}}]}\}$.
A point on a midpoint cross section is then parametrized as:
\begin{align}
\deform(\q, z_{[{j - \onehalf{}}]}) &=
\ct(z_{[{j - \onehalf{}}]}) + s(z_{[j - \onehalf{}]})\rot(z_{[{j - \onehalf{}}]})\q
\\
\ct(z_{[j - \onehalf{}]}) &\equiv \tfrac{1}{2}
\left[ {\ct(z_{[j-1]}) + \ct(z_{[j]})} \right]
\\
s(z_{[j - \onehalf{}]}) &\equiv \tfrac{1}{2}
\left[ {s(z_{[j-1]}) + s(z_{[j]})} \right]
\end{align}
This is similar to the staggered grid discretization of previous work \cite{rods,flexible_2006}, where the frames $\rot$ are stored at the \emph{midpoints}.
Contrary to previous approaches, our model also has a scale that we store at the \emph{endpoints} of the linear elements.
We can now rewrite \Equation{variational_implicit} as:
\begin{equation}
\min_{\{\ct_t, s_t, \rot_t\!\}} \sum_{j=1}^m l_{[j]} \!\!\!\!\!\!
\iint\limits_{D(z_{[j - .5]})}
\!\!\!\!\!
\tfrac{m}{2h^2}\|\deform_{t} - \predict_t\|_2^2
+ 
\tfrac{1}{2}{\textstyle \sum}_i\| \pot_i(\deform_t, \rest) \|_{\stiff_i}^2 dx\,dy
\label{eq:variational_implicit_discretized}
\end{equation}

%% file: fig/static+dynamic/item.tex
\begin{figure*}
\begin{minipage}[t]{.99\columnwidth}
\begin{overpic} 
[width=\columnwidth]
{\currfiledir/litem.pdf}
\end{overpic}
\caption{
\textbf{Static behavior} --
We compare the deformation of a standard cosserat rod to our volumetric invariant version.
Our additional degrees of freedom allow us to model the buckling (resp. bulging) caused by the stretching (resp. compression) of the rod.
}
\label{fig:static}
\end{minipage}%
$\hfill$
\begin{minipage}[t]{.99\columnwidth}
\begin{overpic} 
[width=\columnwidth]
{\currfiledir/ritem.pdf}
\end{overpic}
\caption{
\textbf{Dynamic behavior} --
Our generalized rods do not only capture static volumetric deformations, but the scale's velocity allows us to model volume dynamics.
In this example, note the rod length is unchanged, but our formulation models a volumetric shockwave travelling through the rod.
}
\label{fig:dynamic}
\end{minipage}
\end{figure*}

%% file: sec/5_rodsolver.tex
\section{Elastic Potentials}
\label{sec:elastic}
In this section we detail the elastic potentials used to simulate our volume preserving rods. 

\paragraph{Strain potential}
We define the strain at a midpoint as 
\begin{equation}
\label{eq:strain}
\text{E}_{\text{strain}}(z_{[j - \onehalf]}) =  \iint\limits_{D(z_{[j - \onehalf]})} \|\rot^T\gradient \deform - \atrest{\rot}^T\gradient \rest\|_{\stiff_s}^2 dx\,dy,
\end{equation}
where $\stiff_s =\left[\begin{smallmatrix} k^x_s \vec{e}^x & k^y_s \vec{e}^y  & k^z_s\vec{e}^z \end{smallmatrix}\right]$ is a diagonal matrix of stiffness parameters and $[\vec{e}^x \vec{e}^y \vec{e}^z]$ is the standard basis.
$\gradient \deform$ and $\gradient \rest$ denote the deformation gradients, i.e., the Jacobian matrices of the deformation functions~\cite{Sifakis:2012}. As $\deform$ and $\rest$ map $\mathbb{R}^3$ to $\mathbb{R}^3$, the Jacobian matrices are $3\times3$.  The Jacobians are not rotationally invariant so we rotate them back to the parametrization domain to be able to compare them on a common ground. As explained in \Appendix{elastic} integrating the strain energy \eq{strain} leads to
\begin{align}
\text{E}_{\text{strain}}(z_{[j - \onehalf]}) &= \pi r^2k^z_s \|\gradient_z \ct - \rotw\atrest{\rotw}^T\gradient_z \atrest{\ct} \|^2_2 \label{eq:stretch_z}\\
&+ \pi r^2(k^x_s + k^y_s)(s - \atrest{s})^2 \label{eq:stretch_xy}\\
&+  \tfrac{1}{4}{\pi r^4(k^x_s + k^y_s)}(\gradient_z s - \gradient_z \atrest{s})^2 \label{eq:surface_stretch}\\
&+  \|s\darboux - \atrest{s}\atrest{\darboux}\|^2_{\mat{H}_s} \label{eq:bending_twist}, 
\end{align}
where $\mat{H}_s = \left[\begin{smallmatrix} {\pi r^4k^z_s \vec{e}^x} & {\pi r^4k^z_s\vec{e}^y }  & {\pi r^4(k^x_s + k^y_s)\vec{e}^z}  \end{smallmatrix}\right]$ is the second moment of area of a disc scaled by the stiffness, and the Darboux vector is denoted by $\darboux = [\Omega^u, \Omega^v, \Omega^w]^T$. Note that we retrieved similar energies in previous works \cite{Kugelstadt:2016} augmented by our additional scale degree of freedom.
The energies \eq{stretch_z} and \eq{stretch_xy} respectively measure the stretch along the curve and the cross section, while \eq{surface_stretch} measures the variation of scale across sections, and \eq{bending_twist} measures bending/twisting. Interestingly, \Equation{surface_stretch} can also be interpreted as a measure of surface stretch.

We use an additional energy measuring the second order variation of scale complementing \eq{surface_stretch} with a measure of surface \emph{bending}
\begin{align}
\text{E}_{\text{bending}}(z_{[j - \onehalf]}) = \tfrac{1}{4} {\pi r^4\left(k^x_b + k^y_b\right)}
\left( 
\gradient^2_z s - \gradient^2_z \atrest{s} \right)^2
\end{align}
where $\gradient^2\!=\!\gradient\cdot\gradient\!=\!\Delta$ is the Laplacian operator. Note that this energy can also been seen as an \emph{approximation} of:
\begin{equation}
\iint\limits_{D(z_{[j - \onehalf]})} \|\rot^T\gradient^2 \deform - \atrest{\rot}^T\gradient^2 \rest\|_{\stiff_b}^2 dxdy,
\end{equation}
where $\stiff_b\!=\!\left[\begin{smallmatrix} k^x_b \vec{e}^x & k^y_b \vec{e}^y  & k^z_b\vec{e}^z \end{smallmatrix}\right]$. 
This energy compares the Laplacians of the deformation functions giving a second order measure of the deformation.

\paragraph{Volume potential}
By denoting the determinant with $|\cdot|$, and stiffness by $k_v$, we define the volume preservation at midpoints as:
\begin{equation}
\label{eq:vol}
\text{E}_{\text{vol}}(z_{[j - \onehalf]}) = \iint\limits_{D(z_{[j - \onehalf]})} k_v \left( |\gradient \deform| - |\gradient \rest| \right)^2 dx\,dy  
\end{equation}
Note that the determinant is rotationally invariant, hence it is not necessary to rotate the Jacobians.
As explained in \Appendix{elastic} integrating the volume energy \eq{vol} leads to:
\begin{align}
\text{E}_{\text{vol}}(z_{[j - \onehalf]}) &= \pi r^2 k_v\|s^2\gradient_z \ct - \atrest{s}^2\rotw\atrest{\rotw}^T\gradient_z \atrest{\ct}\|^2_2 \\  
&+ \tfrac{1}{2}{\pi r^4 k_v}\left(s^3\Omega^u - \atrest{s}^3\atrest{\Omega}^u\right)^2 \\ 
&+ \tfrac{1}{2}{\pi r^4 k_v}\left(s^3\Omega^v - \atrest{s}^3\atrest{\Omega}^v\right)^2.
\end{align}

In~\Appendix{appeval}, we include a comparison between the volume preservation of our method and the inter-step update method of \citeN{bergou2010discrete}.

\input{fig/octopus/item.tex}

%% file: fig/octopus/item.tex
\begin{figure*}[t]
\centering
\begin{overpic} 
[width=\linewidth]
{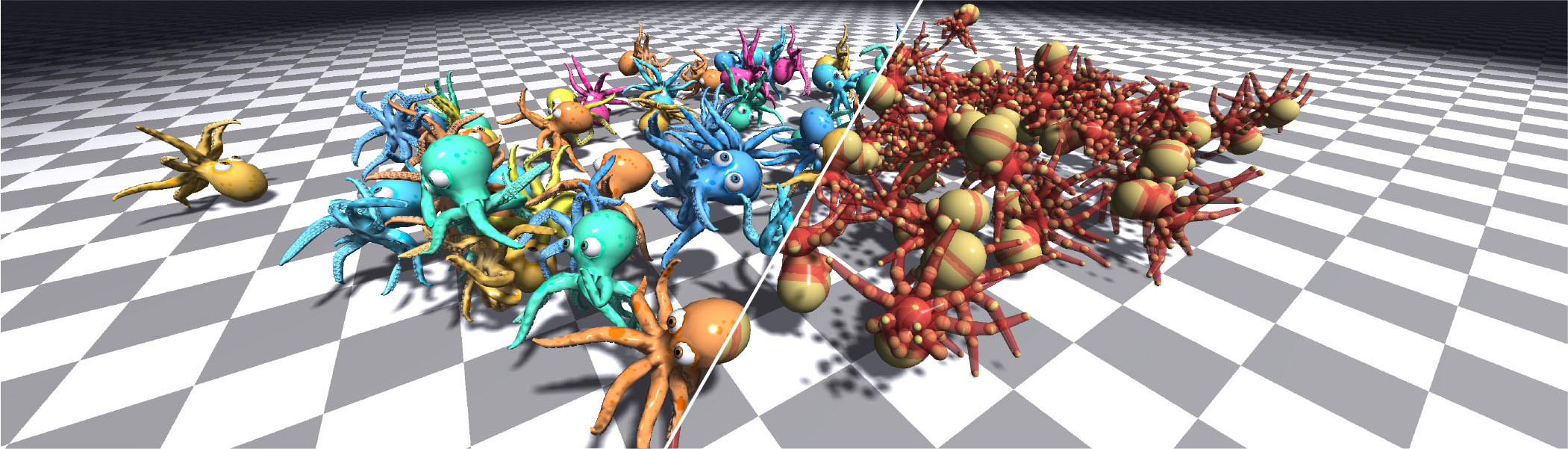}
\end{overpic}
\caption{
Being built on VIPER primitives, our physics engine can simulate soft-body deformation and dynamic interactions between hundreds of models in \emph{real-time}.
A peculiarity of our engine is that \emph{both} collision and simulation are executed on the \emph{same} geometry; 
see our video in the \textbf{additional material}.
}
\label{fig:octopus}
\end{figure*}

%% file: sec/6_optimization.tex
\section{Optimization} 
\label{sec:optimization}

To approach the non-linear optimization problem in~\eq{variational_implicit_discretized}, we linearize the inertia and non-linear elastic terms, and then solve the optimization iteratively in a Gauss-Newton fashion. To warm start the optimization, we first compute a \emph{prediction step} by ignoring the elastic potentials and by solely minimizing the inertia term -- this simply provides an initial guess. We then compute a (set~of)~\emph{correction steps} that also include the elastic potentials.

\paragraph{Prediction step}
By denoting by $\boldsymbol{\theta}$ the angles parametrizing the rotation matrix and $\mathcal{I}$ is the second moment of area of a disc in world-space, the predictions for the different degrees of freedom of our model are computed as:
\begin{align}
\hat{\ct}  &= \ct + h\dot{\ct} + \tfrac{h^2}{\pi r^2 m}\boldsymbol{\xi}_{\text{ext}},
&\text{(center prediction)} \\
\hat{\boldsymbol{\theta}} &= \boldsymbol{\theta} + h \dot{\boldsymbol \theta}  +  \tfrac{\mathcal{I}^{-1}h^2}{s^2m}\boldsymbol{\tau}_{\text{ext}},
&\text{(frame prediction)} \\
\hat{s} &=  s + h\dot{s} + \tfrac{2h^2}{\pi r^4 m}\boldsymbol{\gamma}_{\text{ext}},
&\text{(scale prediction)}
\end{align}
where $\boldsymbol{\xi}_{\text{ext}}$ is the sum of the external forces which act on the disc, $\boldsymbol{\tau}_{\text{ext}} $ is the sum of the external torques, and $\boldsymbol{\gamma}_{\text{ext}}$ is a quantity which can be seen as the counterpart of the total external torque measuring the sum of the external forces projected on the position vectors; see \Appendix{prediction_updates} for more details.
Note that the center and frame predictions are similar to the rigid-body equations of motion for a stretched disc.
On top of these equations, we get a scale prediction describing how the scale of the disc is affected by the velocity and the external forces.

\paragraph{Correction steps} We then compute a set of correction steps including both inertial/elastic terms. We define 
\begin{equation}
\vec{X} = [\ct_{[0]}^T, s_{[0]}, \boldsymbol{\theta}_{[\onehalf]}^T, \ct_{[1]}^T, s_{[1]}, \boldsymbol{\theta}_{[1\onehalf]}^T, \cdots]^T
\end{equation}
as the vector containing all the degrees of freedom, and $\boldsymbol{\lambda}$ as the vector of Lagrange multipliers. $\vec{K}$ is a block diagonal matrix containing the stiffness parameters multiplied by the length of the piecewise elements, $\mat{A}$ is a block diagonal matrix stacking the inertia weights multiplied by the length of the piecewise elements, and 
\begin{equation}
\pot(\vec{X}) = [\pot_1(\vec{X}), \pot_2(\vec{X}), \cdots]^T
\end{equation}
stacks the potential energy functions.
Denoting the iteration number with $k$, the state is then updated as $\vec{X}^k = \vec{X}^{k-1} + \Delta \vec{X}$ and $\boldsymbol{\lambda}^{k} = \boldsymbol{\lambda}^{k-1} + \Delta \boldsymbol{\lambda}$, where, as derived in \Appendix{correction_updates}:
\begin{align}
\Delta \vec{X} &= -h^2\mat{A}^{-1} \nabla \pot(\vec{X}^{k-1})^T \Delta \boldsymbol{\lambda},
\\
\Delta \boldsymbol{\lambda} &\!=\! 
\left( h^2 
\| \nabla\pot(\vec{X}^{k-1})^T \|^2_{\mat{A}^{-1}} \!+\!\stiff^{-1}\right)^{-1}
\pot(\vec{X}^{k-1}) -  \stiff^{-1}\boldsymbol{\lambda}^{k-1}.
\label{eqn:app_dual_formulation_reformated}
\end{align}

For realtime performance we opt for using an iterative linear system solver such as block Jacobi or Gauss-Seidel. The update for the $i$-th constraint is: 
\begin{equation}
\Delta \boldsymbol{\lambda}_i =  \beta\left(h^2
\| {\pot}_i(\vec{X})^T \|^2_{\mat{A}^{-1}}
+ \stiff^{-1}_{i} \right)^{-1}
\left(\pot_i(\vec{X}) -  \stiff^{-1}_{i}\boldsymbol{\lambda}_i \right)
\label{eq:app_dual_formulation_update}
\end{equation}
where we dropped the superscripts to improve readability, and $\beta$ is a relaxation parameter. Note that $\mat{A}$ being a block diagonal matrix with block of size at most $3 \times 3$, $\mat{A}^{-1}$ can be efficiently computed. Note that \Equation{app_dual_formulation_update} is a generalization of the XPBD update~\cite{xpbd} derived for our volume preserving rod model.

%% file: sec/7_engine.tex
\section{Real-time physics engine}
\label{sec:engine}
We implemented our real-time solver on a GPU by leveraging the \emph{Thrust} framework provided by the CUDA library, which we execute on a single NVIDIA GTX 1080 graphics card. 
In our engine, any volumetric object is modeled as a collection of 
tapered capsule primitives, or ``\pill{}s'' (for brevity), while the floor is represented as a simple halfplane constraint.
During simulation, at each time step, we start by first animating kinematic objects (e.g. bones).
We then compute the inertial predictions, and perform collision detection (\Sec{detection}) to generate collision resolution constraints (\Sec{handling}).
We then solve for all constraints using a Jacobi solver: constraints are computed in parallel, and the resulting positional displacements are averaged out by a reduction.
The transformations of VIPERs can then be skinned to any surface mesh model (\Sec{skinning}).

Our model provides a viable alternative to cages as a compact control structure for soft-body deformations.
We demonstrate this in \Figure{octopus}, where we rigged a simple octopus character using rods, and solve for collisions and soft-body deformations in real time.
As shown in the accompanying video, our \emph{non-optimized} prototype achieves real-time performance ($\approx 7 ms$ sim, $\approx 6 ms$ render) for scenes containing up to 100 octopuses, each rigged with 37 \pill{}s, whose deformation is skinned to a triangular surface mesh of $\approx 13k$ faces.
Note how for robust collision detection we need far fewer \pill{}s than the volumetric particles used in~\citeN{flex}. 

\input{fig/band/item.tex}

\subsection{Collision detection}
\label{sec:detection}
To detect collisions between physical primitives, we adopt the approach presented by~\citeN{green2010particle} popularized in the context of GPU particle fluid simulation.
Towards this goal, we approximate each \pill{} by its bounding sphere, and (conservatively) detect collisions to be later handled in the resolution loop detailed in~\Section{handling}.
Collisions are detected with the assistance of a uniform grid with cell width chosen as the diameter of the largest bounding sphere, such that collisions between spheres centered in non-neighboring cells are impossible.
Extending~\cite{green2010particle}, we also count the spheres in the neighboring cells of each particle, and use this information to construct \emph{in parallel} a list of all potential collisions.
This is in contrast to the original algorithm which for each sphere processes neighbors in series, and therefore degrades to a partially serial algorithm in cases where many particles fall into a single grid cell.
Further details regarding this process are provided together with an executable 1D example in the form of a \textit{Jupyter notebook} in our additional material.

\subsection{Collision handling}
\label{sec:handling}
In a generic physics engine one would implement collision detection/response between any pair of available proxies.
For the sake of efficiency, in our framework we only tackle two collision proxies: (kinematic) \emph{half-planes} and (dynamic) \emph{\pill{}s}.
Within the combinatorial set of collision pairs, the main challenge is \pill{}-to-\pill{} collisions.
In what follows, we first compute the meta-parameters of the collisions, that are then resolved in the optimization via PBD constraints~\cite{pbd}. 

\paragraph{Collision metadata}
Given two \pill{}s $\vec{P}_a$ and $\vec{P}_b$, each modeled as a pair of spheres, for example, $\vec{P}_a=\{(\ct^a_1,r^a_1),(\ct^a_2,r^a_2)\}$, the fundamental queries we need to answer are:
\CIRCLE{1}~is there a collision?
\CIRCLE{2}~what is the collision point/normal?
\CIRCLE{3}~what is the inter-penetration amount?
As typical in efficient collision resolution, we introduce a \emph{single} PBD constraint modeling  collision forces corresponding to the \emph{largest} \pill{}-to-\pill{} inter-penetration.
In our solution, we leverage the geometric structure of the problem: 
\CIRCLE{1}~a~\pill{} can be interpreted as the union of infinitely many spheres whose position and radii are linearly interpolated between its endpoints;
\CIRCLE{2}~the largest inter-penetration corresponds to the inter-penetration between any pair of spheres, one in pill $\vec{P}_a$ and one in pill $\vec{P}_b$.
By first defining the LERP operator as $\mathcal{L}(\vec{x}_1,\vec{x}_2,\gamma) \equiv (1-\gamma) \: \vec{x}_1 + \gamma \: \vec{x}_2$, the interpolated sphere $\left(\ct(\gamma), r(\gamma) \right)$ is derived by LERP'ing the endpoint quantities as $\ct(\gamma)=\mathcal{L}(\ct_1,\ct_2,\gamma)$ and $r(\gamma)=\mathcal{L}(r_1,r_2,\gamma)$.
The largest inter-penetration is then given by the solution of the \emph{bivariate} optimization problem:
\begin{equation}
\argmin_{\alpha, \beta} \:\: \| \ct_a(\alpha) - \ct_b(\beta) \|_2 - \left(r_a(\alpha) + r_b(\beta) \right)
\end{equation}
Because a closed-form operator $\Pi_b(\alpha)$ providing the barycentric coordinate of the closest-point projection of a point $\ct_a(\alpha)$ onto the \pill{} $\vec{P}_b$ is available (see \Appendix{projection}) we can further simplify this problem into a \emph{scalar} optimization problem:
\begin{equation}
\argmin_{\alpha} \:\: \| \ct_a(\alpha) - \ct_b(\Pi_b(\alpha)) \|_2 - \left(r_a(\alpha) + r_b(\Pi_b(\alpha)) \right)
\label{eq:collsearch}
\end{equation}
which we solve by Dichotomous Search \cite{Antoniou:2007} with a fixed number of iterations (set to 10 in our engine).

\paragraph{Collision constraints}
Having detected a collision between two \pill{}s $\vec{P}_a$ and $\vec{P}_b$, and having computed $\alpha^*$ (and hence $\beta^*=\Pi_b(\alpha^*)$) by solving \eq{collsearch}, we can express a constraint that correlates radii and positions on the two \pill{}s in order to resolve the collision in a least squares sense:
\begin{align}
E_{\text{collision}} = 
\left ( \| \ct_a(\alpha^*) - \ct_b(\beta^*) \|_2 - r_a(\alpha^*) - r_b(\beta^*) \right )^2.
\end{align}

\subsection{Scale-invariant shape matching -- \emph{Bundling}}
\label{sec:bundling}
To represent more complex geometry than individual rods, such as that in \Figure{fibers}, we can gather a collection of rods in a cross-section, and introduce a constraint to explain their joint deformation.
We employ the assumption that muscle fibers in a muscle cross-section contract isotropically.
Indexing the rods in a cross-section by $i$, our rod deformation model can be expressed as a similarity transform
\begin{equation}
\mat{T}_i = \begin{bmatrix}s_i\rot_i & \ct_i \\ \vec{0} & 1 \end{bmatrix}.
\label{eq:shapematch}
\end{equation}
As illustrated in \Figure{shapematch}, for each muscle cross-section we define a scale invariant shape-matching energy measuring the deviation of the current rod deformations $\vec{T}_i$ from a global similarity transform $\vec{T}^*$ of the rest deformations $\bar{\vec{T}}_i $ as
\begin{equation}
E_{\text{shape}} = \sum_i \| \vec{T}^*\bar{\vec{T}}_i - \vec{T}_i \|_2^2.
\label{eq:shapematch}
\end{equation}
We treat this energy as a hard constraint by finding the optimal $\vec{T}^*$ and setting $ \vec{T}_i =  \vec{T}^*\bar{\vec{T}}_i$ as a post-processing step after few iterations. The optimal  $\vec{T}^*$ can be computed following the derivation in~\cite{umeyama:1991}. The optimal rotation $\rot^*$ can be found by solving
\newcommand{\rowvec}[2]{[#1 \quad #2]}
\begin{equation}
\rot^* = \argmin_{\rot \in \text{SO(3)}} \:\: \lVert \rot - \Sigma_i 
\rowvec{s_i \rot_i}{\ct_i - \boldsymbol{\mu}}
\rowvec{\bar{s}_i \bar{\rot}_i}{\bar{\ct}_i - \bar{\boldsymbol{\mu}}}^T
\rVert_2^2,
\end{equation}
where $\boldsymbol{\mu} = \tfrac{1}{n}\Sigma_i \ct_i$. We compute the optimal rotation $\rot^*$ using the robust approach presented in~\cite{Muller:2016}.
The optimal scale can be computed as
\begin{equation}
s^* =  \frac{\sum_i \matsum(\rot^*
\rowvec{\bar{s}_i \bar{\rot}_i}{\bar{\ct}_i  - \bar{\boldsymbol{\mu}}}
\circ 
\rowvec{s_i\rot_i}{\ct_i - \boldsymbol{\mu}})}
{\sum_i\matsum(
\rowvec{\bar{s}_i \bar{\rot}_i}{\bar{\ct}_i - \bar{\boldsymbol{\mu}}}
\circ
\rowvec{\bar{s}_i\bar{\rot}_i}{\bar{\ct}_i - \bar{\boldsymbol{\mu}}})},
\end{equation}
where $\matsum(\cdot)$ adds all entries of the matrix and $\circ$ is the Hadamard product. Finally, the optimal translation can be derived as
\begin{equation}
\mat{c}^* =  \boldsymbol{\mu} - s^* \rot^* \bar{\boldsymbol{\mu}}.
\end{equation}

\input{fig/shapematch+contract/item}

\subsection{Skinning VIPERs to surface deformation}
\label{sec:skinning}
Our VIPER rods can also be employed as a volumetric \emph{proxy} driving the deformation of a surface mesh; see \Figure{teaser} and \Figure{band}.
In particular, we blend the relative transformations between deformed and rest pose configurations via \emph{linear blend skinning}.
To achieve this, we utilize a modified version of the LBS weight computation by \citeN{sphmsh}.
In that paper, weights were computed by fairing an initial assignment where each vertex was \emph{fully} attached to the nearest element.
Instead, we modify this assignment to begin with weights for each vertex that are proportional to the inverse square-distance from the vertex to the surface of each pill. This resolves cases where multiple surfaces are at equal distance, and the nearest neighbor is multiply-defined.

\subsection{Energy implementation}
The energies derived in Section~\ref{sec:elastic} have been derived using continuous operators. To implement these energies we approximate  $\gradient_z \vec{c}(z_{[j + \onehalf]})$,  $\gradient_z s(z_{[j + \onehalf]})$, and $\gradient^2_z s(z_{[j + \onehalf]})$ using finite difference such that
\begin{align}
\gradient_z \vec{c}(z_{[j + \onehalf]}) &=
l_{[j]}^{-1}
\left(
{\vec{c}(z_{[j+1]}) - \vec{c}(z_{[j]})}
\right),
\\
\gradient_z s(z_{[j + \onehalf]}) &= 
l_{[j]}^{-1}
\left( 
{s(z_{[j+1]}) - s(z_{[j]})}
\right),
\\
\gradient^2_z s(z_{[j]}) &=
l_{[j]}^{-1} \left( {s(z_{[j+1]}) - s(z_{[j]})} \right) 
\\
&- l_{[j-1]}^{-1} \left( {s(z_{[j]}) - s(z_{[j-1]})} \right). \nonumber
\end{align}
For simplicity of the derivations we used angles $\boldsymbol{\theta} $ to parametrize the rotation $\rot$ . However, our implementation uses quaternions. For small angles the corresponding quaternion is $\mathbf{\vec{Q}} = [\tfrac{\boldsymbol{\theta}}{2}^T, 1]^T$. We also use quaternions to represent rotations and approximate the Darboux vector using 
\begin{align}
\darboux(z_{[j]}) = 4 (l_{[j-1]} + l_{[j]})^{-1} \operatorname{Im}(\bar{\mathbf{\vec{Q}}}_{[j-.5]} \mathbf{\vec{Q}}_{[j+.5]}),
\end{align}
where $\operatorname{Im}(\cdot)$ gives the imaginary part of a quaternion.

%% file: fig/band/item.tex
\begin{figure*}[t]
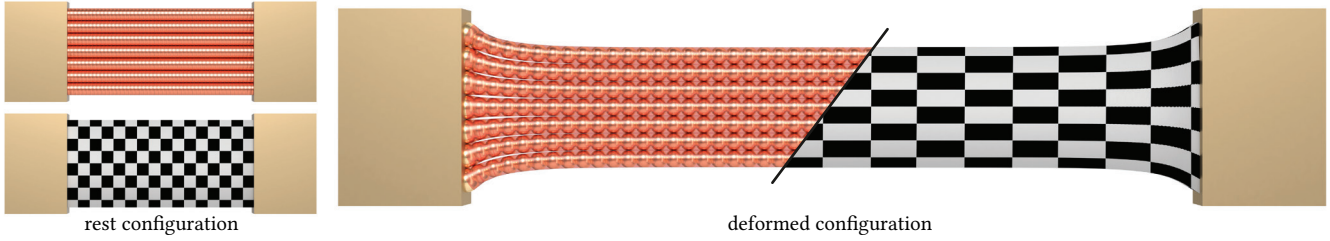

\centering
\begin{overpic} 
[width=\linewidth] 
{\currfiledir/fig.pdf}
\put(6.3,0){\small{rest configuration}}
\put(54,0){\small{deformed configuration}}
\end{overpic}
\caption{
(left) An elastic band mesh and its VIPER discretization.
(right) The VIPER simulation and their deformation ``skinned'' to the rest-pose mesh.
}
\label{fig:band}
\end{figure*}

%% file: fig/shapematch+contract/item.tex
\begin{figure*}
\begin{minipage}[t]{\columnwidth}
\begin{overpic} 
[width=\columnwidth]
{\currfiledir/litem.pdf}
\put(7,18){\footnotesize{rest configuration}}
\put(7.5,15){\footnotesize{(and constraints)}}

\put(93,42){\rotatebox{90}{\footnotesize{traditional}}}
\put(96,39){\rotatebox{90}{\footnotesize{shape-matching}}}

\put(93,10){\rotatebox{90}{\footnotesize{scale-invariant}}}
\put(96,9.3){\rotatebox{90}{\footnotesize{shape-matching}}}
\end{overpic}
\caption{
\textbf{Scale-invariant shape matching} --
Shape matching can recover a rigidly transformed configuration, while our model allows for a null-space that includes uniform scale.
We employ this model as we work under the assumption that muscle fibers in a muscle cross-section contract isotropically.
}
\label{fig:shapematch}
\end{minipage}%
$\hfill$
\begin{minipage}[t]{\columnwidth}
\begin{overpic} 
[width=\columnwidth]
{\currfiledir/ritem.pdf}
\put(93,42.4){\rotatebox{90}{\footnotesize{muscle at rest}}}
\put(96,42.2){\rotatebox{90}{\footnotesize{(not activated)}}}

\put(93,24){\rotatebox{90}{\footnotesize{traditional}}}
\put(96,21){\rotatebox{90}{\footnotesize{shape-matching}}}

\put(93,0){\rotatebox{90}{\footnotesize{scale-invariant}}}
\put(96,-0.5){\rotatebox{90}{\footnotesize{shape-matching}}}
\end{overpic}
\caption{
\textbf{Muscle contraction} -- 
(top) Muscle at rest and its excitation (force shortening edge lengths in an area) with traditional shape-matching constraints (middle) vs our novel scaled shape-matching constraints (bottom).
}
\label{fig:contract}
\end{minipage}
\end{figure*}

%% file: sec/8_muscles.tex
\section{Anatomical modeling and simulation}
\label{sec:anatomical}
We now describe how our rods can be used to model complex anatomical structures such as \emph{bones} and \emph{muscles}; see \Figure{teaser}.
For the former, we use a simplified version of \citeN{sphmsh} where only \pill{} primitives are used -- this primarily allows us to reduce the complexity of the collision detection/resolution codebase.
We then detail the conversion of digital models of muscles into VIPERs in \Section{viperization}, and describe a few nuances about the simulation of their motion in \Section{musclesim}.

\subsection{Muscles to rods conversion -- \emph{Viperization}}
\label{sec:viperization}
\newcommand{\nrods}{K}
\newcommand{\nslices}{M}
We developed a (weakly-assisted) technique to convert a traditional muscle model into a collection of $\nrods{}$ rods.
As outlined in \Figure{fiberizer}, our process involves several phases.
We begin by computing a volumetric discretization of the muscle's surface.
We then ask the artist to paint annotations on the surface marking the start (sources) and the end (sinks) of the muscle.
A harmonic solve like the one described in~\citeN{choi_plos13} is then executed to compute a field that smoothly varies in the $[0, 1]$ range along the muscle's length.
We then sample $\nslices{}$ iso-levels of this field to be surfaces containing the $\nslices{}$ vertices of \emph{each} of the $\nrods{}$ generated VIPERs.
Within each iso-level we require: \CIRCLE{1}~VIPERs to have the same radii, and \CIRCLE{2}~to~be distributed uniformly on the slice (iso-surface).
To obtain this, we perform a \emph{restricted centroidal Voronoi diagram} of $\nrods{}$ points on each slice~\cite{pmp}, while simultaneously penalizing the length of each rod.
We alternate this variational optimization with a discrete one that re-assigns spheres to different rods in order to minimize the sum of rod lengths.
This process, which we refer to as ``combing'', starts from one end of the muscle, and executes in order $\nslices\!-\!1$ instances of minimum-cost bipartite matching (which we solve in polynomial time via the Hungarian algorithm), where the pairwise costs are the $\nrods{}^2$ euclidean distances across rod nodes in two adjacent slices.
A few examples of the results of this process are illustrated in \Figure{teaser} and \Figure{muscles}.

\input{fig/fiberizer+muscles/item}

\subsection{Muscle simulation}
\label{sec:musclesim}
Once a muscle is \emph{viperized} as described in \Section{viperization}, rod centers within the same cross-section are connected via the \emph{bundling} constraints in \Eq{shapematch}.
Every rod also obeys the constraints described in \Section{elastic}.
The endpoints of rods are kinematically attached to bones, and intra-muscle collisions are disabled, while inter-muscle collisions are detected and resolved as described respectively in \Section{detection} and \Section{handling}.
Muscles can also be \emph{activated} (fiber contractions generating a stronger force, producing a change of shape at constant muscle length and volume) by inserting internal forces, or even simply shortening the length of fibers resulting in the bulging effects illustrated in \Figure{contract}.
We also speed-up the solver convergence during fast motion by exploiting the availability of bone transformations.
In more detail, each muscle particle has two skinning weights corresponding to the two bones the muscle is attached to.
At the beginning of each frame we use the transform of each bone relative to their last frame's transforms to initialize the displacement of the particle using LBS, and later refine this via simulation.

%% file: fig/fiberizer+muscles/item.tex
\begin{figure*}
\begin{minipage}[t]{\columnwidth}
\begin{overpic} 
[width=\columnwidth]
{\currfiledir/litem.pdf}
\end{overpic}
\caption{
The mesh-to-VIPER conversion process.
Given source/sink constraints, we compute a harmonic function in the volume, and extract a few discrete iso-levels. Within each of these, we execute a restricted CVD to place 5 elements of the same radii on these surfaces.
We then execute a combinatorial optimization that connects samples across layers to produce minimal length curves.
}
\label{fig:fiberizer}
\end{minipage}%
$\hfill$
\begin{minipage}[t]{\columnwidth}
\begin{overpic} 
[width=\columnwidth]
{\currfiledir/ritem.pdf}
\put(14,0){\small{pectoral}}
\put(50,0){\small{biceps}}
\put(78,0){\small{deltoid}}
\put(97.5,43){\rotatebox{90}{\small{input}}}
\put(97.5,12){\rotatebox{90}{\small{output}}}
\end{overpic}
\caption{ 
We illustrate several examples of the VIPERs extracted by our automated process.
For the \emph{pectoral}, our VIPER model employs 8 rods, each discretized by 8 elements.
In comparison, the surface meshes by ZIVA contain ($|\partial V|=1584, |\partial F|=3164$) on the boundary and its (volumetric) simulation mesh contains ($|V|=350, |T|=1089$) tetrahedral elements. Where $\partial \equiv \text{``boundary''}$, $F \equiv \text{``faces''}$, $V \equiv \text{``vertices''}$ and $T \equiv \text{``tetrahedra''}$.
}
\label{fig:muscles}
\end{minipage}
\end{figure*}

%% file: sec/10_conclusion.tex
\section{Conclusions \& future work}
\label{sec:future}
In this paper, we introduced a novel formulation of cosserat rods that considers local volume, and optimizes for its local conservation.
The resulting position-based simulation is highly efficient, and is relatively straightforward to implement on graphics hardware.
We demonstrated how rod-bundling is a powerful representation for the modeling of volumetric deformation~--~and in particular for skeletal muscles.
Rather than requiring artists to model from
scratch, we also introduced an algorithm to procedurally generate VIPERs with minimal user interaction. 
Finally, by coupling the rod simulation to a surface mesh via skinning, our model can be thought of as a direct alternative to tetrahedral meshes and cages for real-time non-rigid deformation.
Most importantly, our generalized rods formulation opens up a number of venues for future work, which we classify in three broad areas, as elaborated below.

\paragraph{Model generalization}
While in our rods we discretized the skeletal curve with piecewise linear elements, it would be interesting to investigate whether using \emph{continuous} curve models such as splines would be tractable -- from both mathematical, as well as implementation standpoints.
Similarly to~\cite{muller_vriphys11,muller_tog11}, our model could be extended to model \emph{anisotropic} deformations, i.e. both the rest pose and deformed rod could have a non-circular cross section.
Further, while in this paper we treated the modeling of non-circular cross-sections via bundling, the theory of medial axis~\cite{skelstar_eg16} tells us how any shape can be approximated via primitives formed by convex-hulls of three-spheres -- what~\citeN{hmodel} called ``wedges''.
Extending our volume-invariant rods models to volume-invariant \emph{wedges} would provide an elegant generalization of our modeling paradigm.
\paragraph{Anatomical modeling}
As highlighted by our supplementary video, rod primitives can be exploited for efficient approximate modeling and simulation of complex structures.
Nonetheless, the dynamics of the human body are the result of the complex \emph{interplay} between muscle, fat, and deformation of skin.
Enriching our model to also account for these factors would be an interesting extension.
For example, rather than driving the muscle surface via skinning as in \Section{skinning}, one could represent a muscle as a controllable \emph{implicit blend}~\cite{angles_siga17}, approximate fat as an elastic offset between muscles and skin, and \emph{simulate skin} as an elastic surface whose vertices lie on a (potentially) user-controlled iso-level of the implicit function.
Further, while artistic editing of physically driven anatomical systems can be difficult due to the complexity of simulation, our framework could enable \emph{interactive modeling}, similar to what ZBrush/ZSphere currently provides for authoring static geometry.
By extending the works in \cite{hmodel,honline}, an efficient anatomical model for a particular user could also be constructed by fitting to 
RGBD data.

\paragraph{Optimization}
Our solver has not \emph{yet} directly leveraged the multi-resolution structure of rod geometry.
More specifically, the curve parameterization of rods offers a domain over which designing prolongation/restriction operators needed for a \emph{geometric multi-grid} implementation becomes straightforward.
An orthogonal dimension for optimization would be to consider the existence of multi-resolution structures within the \emph{cross-sectional} domain; see \Figure{fibers}.
This could be exploited in offering multi-scale interaction for artists in editing our rod models, as well as producing LOD models for efficient simulation.
Finally, while we employed out-of-the-box geometry processing tools to convert a triangular mesh into a rod model, we believe fitting a fiber-bundle model to a given solid could be achieved without the (often finicky) conversion to tet mesh, but rather as a direct optimization over fiber placements.

%% file: sec/12_appendix_energies.tex
\section{Elastic Potentials}
\label{app:elastic}
\paragraph{Strain} As defined in \Section{elastic} we can write the strain energy as
\begin{equation}
\label{eq:app:strain}
\text{E}_{\text{strain}} =  \iint\limits_{D} \|\rot^T\gradient \deform - \atrest{\rot}^T\gradient \rest\|_{\stiff}^2 dx dy,
\end{equation}
where we drop the subscript from $\stiff$ for brevity. This equation can be extended as
\begin{equation}
\iint\limits_{D} \|\rot^T \gradient(\ct + s\rot \q) - \atrest{\rot}^T \gradient(\atrest{\ct} + \atrest{s} \atrest{\rot} \q) \|^2_{\mat{K}} dxdy,
\end{equation}
where
\begin{align}
\rot^T\gradient(\ct + s\rot \q) &= \rot^T(\gradient \ct + \gradient s \rot \q +  s\gradient \rot \q +  s\rot \gradient \q), \\
&= \rot^T\gradient \ct + \gradient s \q +  s\rot^T\gradient \rot\q +  s\gradient \q.
\end{align}
We can derive the gradient operator for each part of this summation leading to
\begin{equation}
\rot^T\gradient \ct = \begin{bmatrix} \vec{0} & \vec{0} & \rot^T \gradient_z \ct  \end{bmatrix} = \left[\begin{smallmatrix}0&0&0\\0&0&0\\0&0&\rotw^T \gradient_z \ct \end{smallmatrix}\right],
\end{equation}
\begin{equation}
\gradient s \q  = \begin{bmatrix} \vec{0} & \vec{0} & \gradient_z s \q \end{bmatrix} = \left[\begin{smallmatrix}0&0&\gradient_z sx\\0&0&\gradient_z sy\\0&0&0\end{smallmatrix}\right],
\end{equation}
\begin{equation}
s\rot^T\gradient \rot\q = \begin{bmatrix} \vec{0} & \vec{0} & s\darboux {\times} \q \end{bmatrix} =  \left[\begin{smallmatrix}0&0&-s\Omega^wy\\0&0&s\Omega^wx\\0&0&s\Omega^uy - s\Omega^vx\end{smallmatrix}\right],
\end{equation} 
\begin{equation}
s\gradient \q  =  \begin{bmatrix} s\vec{e}^x & s\vec{e}^y & \vec{0} \end{bmatrix} =  \left[\begin{smallmatrix}s&0&0\\0&s&0\\0&0&0\end{smallmatrix}\right],
\end{equation}
where $\darboux = [\Omega^u \Omega^v \Omega^w]^T$ is the Darboux vector; and analogous expressions can be derived for $\gradient \rest$.
We can now observe that \eq{app:strain} can be broken up into the sum of distinct energies
\begin{align}
\label{eq:app:strain_simplified}
\text{E}_{\text{strain}} =  \iint\limits_{D}
& k^z (\rotw\at{^T} \gradient_z \ct - \atrest{\rotw}^T\gradient_z \atrest{\ct})^2 \\
+&  \|(\gradient_z s - \gradient_z \atrest{s})\q\|^2_{\mat{K}}\\
+& \|(s\darboux - \atrest{s}\atrest{\darboux}) \times \q\|^2_{\mat{K}}\\
+& (k^x + k^y)(s - \atrest{s})^2 dx dy,
\end{align}
as the cross terms evaluate to $\vec{0}$.
After integrating over the disc we can reformulate \eq{app:strain_simplified} as
\begin{align}
\text{E}_{\text{strain}} &= \pi r^2k^z \|\gradient_z \ct - \rotw\atrest{\rotw}^T\gradient_z \atrest{\ct} \|^2_2 \\
&+  \tfrac{\pi r^4(k^x + k^y)}{4}(\gradient_z s - \gradient_z \atrest{s})^2\\
&+  \|s\darboux - \atrest{s}\atrest{\darboux}\|^2_{\mat{H}}\\
&+ \pi r^2(k^x + k^y)(s - \atrest{s})^2,
\end{align}
where $\mat{H} = \left[\begin{smallmatrix} \frac{\pi r^4k^z \vec{e}^x}{4} & \frac{\pi r^4k^z\vec{e}^y }{4}  & \frac{\pi r^4(k^x + k^y)\vec{e}^z}{4}  \end{smallmatrix}\right]$ is the second moment of the area of a disc scaled by the stiffness.

\paragraph{Volume} As defined in \Section{elastic} we can write the volume energy as
\begin{align}
\text{E}_{\text{vol}} &= \iint\limits_{D} k(|\gradient \deform| - |\gradient \rest|)^2 dxdy  \\
&= \iint\limits_{D} k(|\rot^T \gradient \deform| - |\atrest{\rot}^T \gradient \rest|)^2 dxdy,
\end{align}
Based on the strain derivation, the determinant can be computed as
\begin{align}
|\rot^T\gradient(\ct + s\rot\q)| &= |\left[\begin{smallmatrix}s&0&\gradient_z sx-s\Omega^wy\\0&s&\gradient_z sy+s\Omega^wx\\0&0&\rotw^T \gradient_z \ct + s\Omega^uy - s\Omega^vx\end{smallmatrix}\right]| \\
&= s^2\rotw^T\gradient_z \ct + s^3(\Omega^uy - \Omega^vx).
\end{align}
We can now integrate over the disc leading to
\begin{align}
\text{E}_{\text{vol}} &= \pi r^2 k\|s^2\gradient_z \ct - s_{\rest}^2\rotw\atrest{\rotw}^T\gradient_z \atrest{\ct}\|^2_2 \\  
&+ \tfrac{\pi r^4 k}{2}(s^3\Omega^u - \atrest{s}^3\atrest{\Omega}^u)^2 \\ 
&+ \tfrac{\pi r^4 k}{2}(s^3\Omega^v - \atrest{s}^3\atrest{\Omega}^v)^2.
\end{align}

%% file: sec/11_appendix.tex
\newcommand{\qbar}{\bar{\vec{q}}}
\newcommand{\intdisk}[1]{\intD #1 d\qbar}
\newcommand{\skewsym}[1]{[#1]_{\times}}
\newcommand{\fext}{f_{ext}(\bar{\vec{X}})}

\section{Prediction Step}
\label{app:prediction_updates}

As described in \eq{variational_implicit_discretized}, the inertial potential over the disc is of the form
\begin{equation}
\label{eq:app:inertia}
\text{E}_{\text{inertia}} =  \iint\limits_{D} \tfrac{m}{2h^2}\|\deform_t - \predict_t\|^2_2dxdy.
\end{equation}
We aim at finding the unknowns $\x_t = \left[\begin{smallmatrix} \ct_t^T & s_t & \boldsymbol{\theta}_t^T \end{smallmatrix} \right]^T$ that minimize \eq{app:inertia} giving us the prediction update. We denote by $\boldsymbol{\theta}$ the angles parametrizing the rotation matrix. Because the deformation function $\deform_t$ is non linear, i.e., due to the rotational degrees of freedom, we rely on a Gauss-Newton iterative scheme for the minimization. We linearize \eq{app:inertia} at $\x^k_t$ leading to
\begin{equation}
\label{eq:app:inertia_minimization}
\argmin_{\dx^k_t} \iint\limits_{D} \tfrac{m}{2h^2}\|\mat{A}\dx^k_t - \vec{b}\|^2_2dxdy,
\end{equation}
where k is the iteration number, and $\dx_t = \left[\begin{smallmatrix}\Delta \ct_t^T & \Delta s_t & \Delta \boldsymbol{\theta}_t^T \end{smallmatrix}\right]^T$. At each iteration we minimize \eq{app:inertia_minimization}, and then apply the update  $\x^{k+1}_t = \x^{k}_t + \dx^k_t$, where we initialize $\x^0_t = \x_{t-1}$. The matrix $\mat{A}$ can be written as $\mat{A} =  \left[\begin{smallmatrix}\vec{I}_{3\times3} & \rot^k_t\q & -s_t^k[\rot_t^k\q]_{\times}\end{smallmatrix}\right]$,
where $[\cdot]_{\times}$ is a cross product skew-symmetric matrix. The vector $\vec{b}$ is defined as $\vec{b} = \p_{t-1} + h\dot{\deform}_{t-1} + \tfrac{h^2}{m}\vec{f}_{\text{ext}} - \p^k_t$ . To compute the prediction updates we will use a single iteration of Gauss-Newton so $\vec{b} = h\dot{\deform}_{t-1} + \tfrac{h^2}{m}\vec{f}_{\text{ext}}$ as $\p^0_t =  \p_{t-1}$. We will now drop the superscripts and the subscripts to improve readability. As \Equation{app:inertia_minimization} is quadratic we can be find its minimum by solving the normal equation
\begin{equation}
\label{eq:app:normal_equation_prediction}
\left(\iint\limits_{D} \tfrac{m}{h^2}\mat{A}^T\mat{A}dxdy\right)\Delta\vec{x} = \iint\limits_{D} \tfrac{m}{2h^2}\mat{A}^T\vec
{b}dxdy.
\end{equation}
Interestingly, the left hand side $\iint\limits_{D} \tfrac{m}{2h^2}\mat{A}^T\mat{A}dxdy$ can be simplified to a block diagonal matrix of the form
\begin{equation}
\left[\begin{smallmatrix} \iint\limits_{D} \tfrac{m}{2h^2} \mat{I}_{3 \times 3} dxdy & \mat{0} &  \mat{0} \\ 
\mat{0} & \iint\limits_{D} \tfrac{m}{2h^2}(\rot\q)^T(\rot\q) dxdy & \mat{0}
\\
\mat{0} & \mat{0} & \iint\limits_{D} \tfrac{ms^2}{2h^2}[\rot\vec{q}]_{\times}^T[\rot\vec{q}]_{\times}dxdy
\end{smallmatrix}\right],
\end{equation}
by noticing that $[\rot\vec{q}]_{\times}^T(\rot\q) = \vec{0}$. Moreover, as the center of mass of the disc is placed at the origin $\iint\limits_{D}m\rot\q dxdy = \vec{0}$ and $ \iint\limits_{D} m[\rot\q]_{\times} = \vec{0}$. We can now integrate the diagonal elements leading to 
\begin{equation}
\begin{bmatrix} \tfrac{\pi r^2m}{h^2} \mat{I}_{3 \times 3} & \mat{0} &  \mat{0} \\ 
\mat{0} & \tfrac{\pi r^4 m}{2h^2} & \mat{0}
 \\
 \mat{0} & \mat{0} & \tfrac{ms^2}{h^2}\mat{\mathcal{I}}
 \end{bmatrix},
 \end{equation}
where $\mat{\mathcal{I}} = \rot\left[\begin{smallmatrix}\tfrac{\pi r^4}{4} \vec{e}^x & \tfrac{\pi r^4}{4} \vec{e}^y & \tfrac{\pi r^4}{2} \vec{e}^z \end{smallmatrix} \right]\rot^T$ is the second moment of area of a disc in world-space. The right hand side $\iint\limits_{D} \tfrac{m}{h^2}\mat{A}^T\vec{b}dxdy$ can be simplified as
\begin{equation}
\begin{bmatrix}
\tfrac{\pi r^2m}{h}\dot{\ct} + \boldsymbol{\xi}_{\text{ext}} 
\\
\tfrac{\pi r^4 m}{2h}\dot{s} + \boldsymbol{\tau}_{\text{ext}}

\\
\tfrac{ms^2\mat{\mathcal{I}}}{h} \dot{\boldsymbol \theta} +  \boldsymbol{\gamma}_{\text{ext}} 
\end{bmatrix},
\end{equation}

where $\boldsymbol{\xi}_{\text{ext}} = \iint\limits_{D}\vec{f}_{\text{ext}}dxdy$ is the sum of the external forces which act on the disc, $\boldsymbol{\tau}_{\text{ext}} = s \iint\limits_{D}(\rot \q) \times \vec{f}_{\text{ext}}dxdy$ is the sum of the external torques and $\boldsymbol{\gamma}_{\text{ext}} = \iint\limits_{D}(\rot \q) \cdot \vec{f}_{\text{ext}}dxdy$ is a quantity which can be seen as the counterpart of the external torque by measuring the external forces applied along the center direction. 

\paragraph{Center update} By solving the linear system \eq{app:normal_equation_prediction} for $\Delta \ct$ we find the prediction update for the center
\begin{equation}
\Delta \vec{c} =  h\dot{\vec{c}} + \tfrac{h^2}{\pi r^2 m}\boldsymbol{\xi}_{\text{ext}},
\end{equation}
As we integrate on a disc located at a midpoint this update is valid for the center of the disc located at this point. We approximate the update over the end points by using the same update rule.

\paragraph{Scale update} Similarly, the scale update can be computed solving the linear system  for $\Delta s$ leading to
\begin{equation}
\label{eqn:update_scale}
\Delta \vec{s} =  h\dot{s} + \tfrac{2h^2}{\pi r^4 m}\boldsymbol{\gamma}_{\text{ext}},
\end{equation}
We also approximate the update over the end points using the same update rule
 
\paragraph{Frame update} The frame update can be computed by solving the linear system  for $\Delta \boldsymbol{\theta}$ leading to
\begin{equation}
\label{eqn:update_angles}
\Delta \boldsymbol{\theta} =  h  \dot{\boldsymbol \theta} +  \tfrac{\mathcal{I}^{-1}h^2}{s^2m}\boldsymbol{\tau}_{\text{ext}}.
 \end{equation}
\section{Correction Step}
\label{app:correction_updates}

\paragraph{Inertia approximation} From the derivation in \Appendix{prediction_updates} we can obtain an approximation of the inertia term as
\begin{equation}
\text{E}_{\text{inertia}} \approx \tfrac{\pi r^2 m}{2h^2} \|\ct_t  - \hat{\ct}_t\|_2^2 +  \tfrac{\pi r^4m}{4h^2}(s_t  - \hat{s}_t)^2 + \tfrac{s^2m}{2h^2}\|\boldsymbol{\theta}_t  - \hat{\boldsymbol{\theta}}_t\|_\mat{\mathcal{I}}^2,
\end{equation}
where
\begin{align}
\hat{\ct}_t\ = \ct_{t-1} + h\dot{\ct}_{t-1} + \tfrac{h^2}{\pi r^2 m}\boldsymbol{\xi}_{\text{ext}}, \\
\hat{s}_t = s_{t-1} + h\dot{s}_{t-1} + \tfrac{2h^2}{\pi r^4 m}\boldsymbol{\gamma}_{\text{ext}}, \\
\hat{\boldsymbol{\theta}}_t = \boldsymbol{\theta}_{t-1} + h\dot{\boldsymbol{\theta}}_{t-1} + \tfrac{\mat{\mathcal{I}}^{-1}h^2}{s^2m}\boldsymbol{\tau}_{\text{ext}},
\end{align}
are the inertial predictions for the different degrees of freedom. The variational form of implicit Euler \eq{variational_implicit_discretized} can then be written in the form
\begin{equation}
\label{eqn:app_variational_implicit_discretized}
\min_{\vec{X}} \tfrac{1}{2h^2}\|\vec{X} - \hat{\vec{X}}\|_\mat{A}^2 + \tfrac{1}{2}\| \pot(\vec{X}) \|_{\stiff}^2,
\end{equation}
where $\vec{X} = \left[\begin{smallmatrix}\ct_{[0]}^T, s_{[0]}, \boldsymbol{\theta}_{[\onehalf]}^T, \ct_{[1]}^T, s_{[1]}, \boldsymbol{\theta}_{[1\onehalf]}^T, \cdots\end{smallmatrix}\right]^T$ and $\hat{\vec{X}}$ are vectors containing all the degree of freedoms and their predictions, $\vec{K}$ is a block diagonal matrix stacking the stiffness parameters scaled by the length of the piecewise elements, $\mat{A}$ is a block diagonal matrix stacking the inertia weights scaled by the length of the piecewise elements, and $\pot(\vec{X}) = \left[\begin{smallmatrix}\pot_1(\vec{X}) & \pot_2(\vec{X}) & \hdots \end{smallmatrix}\right]^T$ stacks the potential energy functions. 

\paragraph{Variational Solver} To solve this optimization we can linearize the elastic potentials and write an iterative Gauss-Newton optimization
\begin{equation}
\label{eqn:app_variational_implicit_iteration}
\min_{\Delta \vec{X}} \tfrac{1}{2h^2}\|\vec{X}^{k-1} + \Delta \vec{X} - \hat{\vec{X}}\|_\mat{A}^2 + \tfrac{1}{2}\|\pot(\vec{X}^{k-1}) + \nabla \pot(\vec{X}^{k-1})\Delta \vec{X} \|_{\stiff}^2,
\end{equation}
where $k$ is the iteration number, $\vec{X}^{k} = \vec{X}^{k-1} + \Delta \vec{X}$, and we initialize $\vec{X}^0 = \hat{\vec{X}}$.
Since Equation~\ref{eqn:app_variational_implicit_iteration} is quadratic in the unknown $\Delta \vec{X}$, we can minimize it with a single linear solve
\begin{equation*}
\label{eqn:app_normal_equation}
\tfrac{\mat{A}}{h^2}(\vec{X}^{k-1} + \Delta \vec{X} -\hat{\vec{X}}) + \nabla \pot(\vec{X}^{k-1})^T\stiff\left(\pot(\vec{X}^{k-1}) + \nabla \pot(\vec{X}^{k-1})\Delta \vec{X}\right)\!=\!\vec{0}.
\end{equation*}
However, the conditioning of this linear system is greatly dependent on how stiff are the elastic potentials. Following the optimization trick presented in~\cite{gould1986accurate}, for elastic potentials with large stiffness a better option is to split the equation above as
\begin{numcases}{}
\tfrac{\mat{A}}{h^2}(\vec{X}^{k-1} + \Delta \vec{X} - \hat{\vec{X}}) + \nabla \pot(\vec{X}^{k-1})^T\boldsymbol{\lambda}^{k} = \vec{0}, \label{eqn:app_dual_formulation1} \\
\stiff^{-1}\boldsymbol{\lambda}^{k} = \pot(\vec{X}^{k-1}) + \nabla \pot(\vec{X}^{k-1})\Delta \vec{X}. \label{eqn:app_dual_formulation2}
\end{numcases}
Note that when the elastic potentials are infinitively stiff $\stiff^{-1}$ vanishes $\boldsymbol{\lambda} = [\lambda_1, \hdots, \lambda_n]^T$ becomes the vector of Lagrange multipliers.  We can now reformulate (\ref{eqn:app_dual_formulation1}) as
\begin{align}
\Delta \vec{X} &= -h^2\mat{A}^{-1} \nabla \pot(\vec{X}^{k-1})^T \boldsymbol{\lambda}^{k} - (\vec{X}^{k-1} - \hat{\vec{X}}) \label{eqn:app_deltax_1} \\
&\approx -h^2\mat{A}^{-1} \nabla \pot(\vec{X}^{k-1})^T \Delta \boldsymbol{\lambda}  \label{eqn:app_deltax_2},
\end{align}
by assuming $\nabla \pot(\vec{X}^{k}) \approx \nabla \pot(\vec{X}^{k-1})$, and where $\boldsymbol{\lambda}^{k} = \boldsymbol{\lambda}^{k-1} + \Delta \boldsymbol{\lambda}$. We initialize $\boldsymbol{\lambda}^{0} = \vec{0}$. This can be proven by induction knowing that $\vec{X}^0 =  \hat{\vec{X}}$ and $\boldsymbol{\lambda}^{0} = \vec{0}$.
By substituting $\Delta \vec{X}$ in the Equation~\ref{eqn:app_dual_formulation2}, we can rewrite the system of equations as
\begin{equation*}
\label{eqn:app_dual_formulation_reformated}
\begin{cases}
\Delta \vec{X} = -h^2\mat{A}^{-1} \nabla \pot(\vec{X}^{k-1})^T \Delta \boldsymbol{\lambda}, \\
(h^2 \|\nabla \pot(\vec{X}^{k-1})^T\|^2_{\mat{A}^{-1}} + \stiff^{-1})\Delta \boldsymbol{\lambda} = \pot(\vec{X}^{k-1}) -  \stiff^{-1}\boldsymbol{\lambda}^{k-1}.
\end{cases}
\end{equation*}
Therefore, $\Delta \vec{X}$ and $\Delta \boldsymbol{\lambda}$ can be found with a single linear solve.

\section{Closed form \pill{} projection}
\label{app:projection}
Given a \pill{} $\vec{P}=\{(\ct_1,r_1),(\ct_2,r_2)\}$, we can compute the closest point distance of a point $\vec{x}$ onto $\vec{P}$ in closed form as
\begin{equation}
d = \|\vec{x} - \ct_1 + t\hat{\vec{j}}l\|_2 - ((1-t)r_1 + t r_2)
\end{equation}
where $l = \|\ct_1 - \ct_2\|_2$ is the pill length, $\hat{\vec{j}} = l^{-1}(\ct_1 - \ct_2)$ is the pill versor, $\vec{p}_s = \ct_1 - \hat{\vec{j}}\left((\vec{x} - \ct_1) \cdot \hat{\vec{j}}\right)$ the orthogonal projection onto the pill segment, $\theta = \arcsin(l^{-1}(r_1 - r_2))$ is the pill slope angle, and $t = \min(\max(-l^{-1}(\vec{p}_s + o\hat{\vec{j}} - \ct_1) \cdot \hat{\vec{j}}, 0), 1)$ is the barycentric coordinate of the projection, where $o = \|\vec{x} - \vec{p}_s\|_2^2 \tan(\theta)$.

%% file: sec/13_appendix_eval.tex
\section{Further evaluations}
\label{app:appeval}

\input{fig/volumestretch/item.tex}
\paragraph{Evaluation of volume conservation -- \Figure{volumestretch}}
We show the relative error of stretching and volume preservation constraints over the sequence shown in the accompanying video.
The stretching constraint is unavoidably violated as the total length of the segment increases, but the volume constraint remains close to satisfied as the radii are reduced. Perfect satisfaction is not possible as we enforce constraints via penalties; our next experiment delves deeper into this observation.

\input{fig/convergence/item.tex}
\paragraph{Evaluation of solver convergence -- \Figure{convergence}}
We show the convergence properties of our solver by constructing a test case with a difficult initial condition (i.e. far from the optima). More specifically, the side bars are \emph{instantaneously} moved to stretch the segments to twice their rest length (at convergence). This is difficult because the violation of constraints must propagate through the entire chain for the solve to converge. This figure also illustrates that our solver currently achieves \emph{linear} convergence properties (line with slope one in a log-log plot). 
The strain energy saturates to a value which would correspond to the peak in \Figure{volumestretch}, while the volume energy cannot reach zero as the solver seeks a pareto optimal solution that balances the two energy terms. 

\input{fig/bergou/item.tex}
\paragraph{Comparison to \cite{bergou2010discrete} -- \Figure{bergou}}
To illustrate the novel properties of our rod formulation, we compare to the method by \citeN{bergou2010discrete}.
This method achieves volume preservation by updating radii \emph{after} every simulation step.
In contrast, as we include them as variables in the simulation, we are able to model dynamics related to radius changes (e.g. volume shockwaves).
Conversely, \cite{bergou2010discrete} simply enforces volumetric preservation, resulting in a significantly dampened simulation.
Further, the method of \citeN{bergou2010discrete} would be \emph{incapable} of modelling a force which directly induces change in radius, such as a rod shrinking and stretching when pinched between two surfaces.
\input{fig/fem_band/item.tex}
\paragraph{Comparison to the Finite Element Method -- \Figure{fem}}
We compare the VIPER discretization to a tetrahedral FEM discretization with the hyperelastic NeoHookean model of Smith et al. \shortcite{smith2018stable}. As the two methods have different parameterizations of the material properties, we chose values for the FEM sheet that produce similar behaviour to the VIPER sheet: For the FEM sheet we use a Young's modulus of $Y=10^6 Pa$ and Poisson's ratio of $\nu=0.49$. Our method shows a close correspondence with the FEM model, with strongly volume preserving deformation.

%% file: fig/volumestretch/item.tex
\begin{figure}[t]
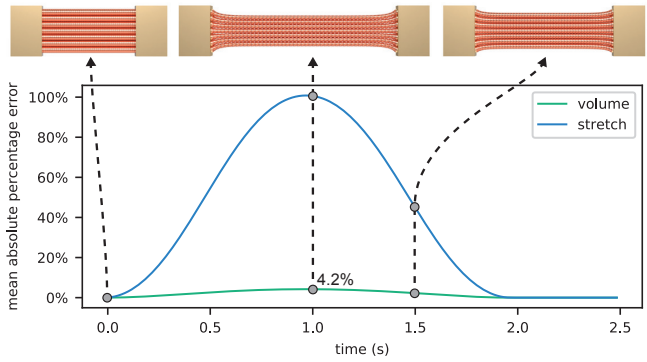

\centering
\begin{overpic} 
[width=\linewidth]
{\currfiledir/fig.pdf}
\end{overpic}
\caption{
Comparison of volume and stretch constraint violations. See the accompanying video in the supplemental materials.
}
\label{fig:volumestretch}
\end{figure}

%% file: fig/convergence/item.tex
\begin{figure}[b]
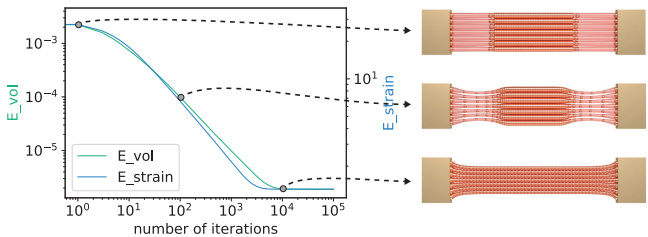

\centering
\begin{overpic} 
[width=\linewidth]
{\currfiledir/fig.pdf}
\end{overpic}
\caption{
Given a difficult initialization, we study the convergence of strain and volume energies in our solver.
}
\label{fig:convergence}
\end{figure}

%% file: fig/bergou/item.tex
\begin{figure}[t]
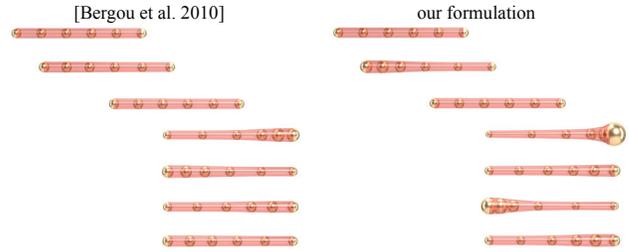

\centering
\begin{overpic} 
[width=\linewidth]
{\currfiledir/fig.pdf}
\end{overpic}
\caption{
The dynamics in our model vs. the solution of~\cite{bergou2010discrete} that simply computes scale in a post-processing stage. See the accompanying video in the supplemental materials.
}
\label{fig:bergou}
\end{figure}

%% file: fig/fem_band/item.tex
\begin{figure}[b]
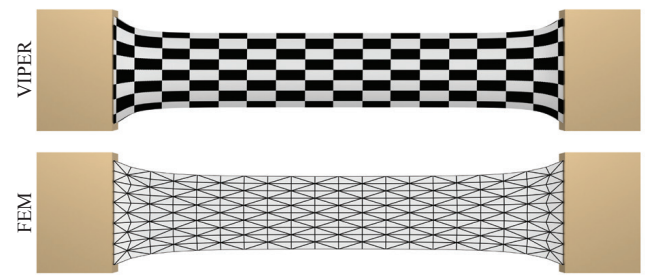

\centering
\begin{overpic} 
[width=\linewidth]
{\currfiledir/fig.pdf}
\end{overpic}
\caption{The volume-preserving behaviour of the VIPER model (upper) vs. the same band simulated with a tetrahedral FEM model (lower).}
\label{fig:fem}
\end{figure}

%% file: paper.bbl
%%% -*-BibTeX-*-
%%% Do NOT edit. File created by BibTeX with style
%%% ACM-Reference-Format-Journals [18-Jan-2012].

\begin{thebibliography}{65}

%%% ====================================================================
%%% NOTE TO THE USER: you can override these defaults by providing
%%% customized versions of any of these macros before the \bibliography
%%% command.  Each of them MUST provide its own final punctuation,
%%% except for \shownote{}, \showDOI{}, and \showURL{}.  The latter two
%%% do not use final punctuation, in order to avoid confusing it with
%%% the Web address.
%%%
%%% To suppress output of a particular field, define its macro to expand
%%% to an empty string, or better, \unskip, like this:
%%%
%%% \newcommand{\showDOI}[1]{\unskip}   % LaTeX syntax
%%%
%%% \def \showDOI #1{\unskip}           % plain TeX syntax
%%%
%%% ====================================================================

\ifx \showCODEN    \undefined \def \showCODEN     #1{\unskip}     \fi
\ifx \showDOI      \undefined \def \showDOI       #1{#1}\fi
\ifx \showISBNx    \undefined \def \showISBNx     #1{\unskip}     \fi
\ifx \showISBNxiii \undefined \def \showISBNxiii  #1{\unskip}     \fi
\ifx \showISSN     \undefined \def \showISSN      #1{\unskip}     \fi
\ifx \showLCCN     \undefined \def \showLCCN      #1{\unskip}     \fi
\ifx \shownote     \undefined \def \shownote      #1{#1}          \fi
\ifx \showarticletitle \undefined \def \showarticletitle #1{#1}   \fi
\ifx \showURL      \undefined \def \showURL       {\relax}        \fi
% The following commands are used for tagged output and should be
% invisible to TeX
\providecommand\bibfield[2]{#2}
\providecommand\bibinfo[2]{#2}
\providecommand\natexlab[1]{#1}
\providecommand\showeprint[2][]{arXiv:#2}

\bibitem[\protect\citeauthoryear{Ali, Liu, Gilles, Kavan, Faure, Palombi, and
  Cani}{Ali et~al\mbox{.}}{2013}]%
        {ali_tog13}
\bibfield{author}{\bibinfo{person}{Dicko~Hamadi Ali}, \bibinfo{person}{Tiantian
  Liu}, \bibinfo{person}{Benjamin Gilles}, \bibinfo{person}{Ladislav Kavan},
  \bibinfo{person}{Fran{\c{c}}ois Faure}, \bibinfo{person}{Olivier Palombi},
  {and} \bibinfo{person}{Marie-Paule Cani}.} \bibinfo{year}{2013}\natexlab{}.
\newblock \showarticletitle{Anatomy transfer}.
\newblock \bibinfo{journal}{\emph{ACM TOG}} (\bibinfo{year}{2013}).
\newblock


\bibitem[\protect\citeauthoryear{Angles, Tarini, Barthe, Wyvill, and
  Tagliasacchi}{Angles et~al\mbox{.}}{2017}]%
        {angles_siga17}
\bibfield{author}{\bibinfo{person}{Baptiste Angles}, \bibinfo{person}{Marco
  Tarini}, \bibinfo{person}{Loic Barthe}, \bibinfo{person}{Brian Wyvill}, {and}
  \bibinfo{person}{Andrea Tagliasacchi}.} \bibinfo{year}{2017}\natexlab{}.
\newblock \showarticletitle{Sketch-Based Implicit Blending}.
\newblock \bibinfo{journal}{\emph{ACM TOG (Proc. SIGGRAPH Asia)}}
  (\bibinfo{year}{2017}).
\newblock


\bibitem[\protect\citeauthoryear{Antoniou and Lu}{Antoniou and Lu}{2007}]%
        {Antoniou:2007}
\bibfield{author}{\bibinfo{person}{Andreas Antoniou} {and}
  \bibinfo{person}{Wu-Sheng Lu}.} \bibinfo{year}{2007}\natexlab{}.
\newblock \bibinfo{booktitle}{\emph{Practical Optimization: Algorithms and
  Engineering Applications}}.
\newblock


\bibitem[\protect\citeauthoryear{Barbi\v{c} and James}{Barbi\v{c} and
  James}{2005}]%
        {Barbic:2005}
\bibfield{author}{\bibinfo{person}{Jernej Barbi\v{c}} {and}
  \bibinfo{person}{Doug~L. James}.} \bibinfo{year}{2005}\natexlab{}.
\newblock \showarticletitle{Real-Time Subspace Integration for St.
  Venant-Kirchhoff Deformable Models}.
\newblock \bibinfo{journal}{\emph{ACM TOG}} (\bibinfo{year}{2005}).
\newblock


\bibitem[\protect\citeauthoryear{Bender, M{\"u}ller, and Macklin}{Bender
  et~al\mbox{.}}{2015}]%
        {pbdstar}
\bibfield{author}{\bibinfo{person}{Jan Bender}, \bibinfo{person}{Matthias
  M{\"u}ller}, {and} \bibinfo{person}{Miles Macklin}.}
  \bibinfo{year}{2015}\natexlab{}.
\newblock \showarticletitle{Position-Based Simulation Methods in Computer
  Graphics.}. In \bibinfo{booktitle}{\emph{Proc. Eurographics (Technical Course
  Notes)}}.
\newblock


\bibitem[\protect\citeauthoryear{Bergou, Audoly, Vouga, Wardetzky, and
  Grinspun}{Bergou et~al\mbox{.}}{2010}]%
        {bergou2010discrete}
\bibfield{author}{\bibinfo{person}{Mikl{\'o}s Bergou}, \bibinfo{person}{Basile
  Audoly}, \bibinfo{person}{Etienne Vouga}, \bibinfo{person}{Max Wardetzky},
  {and} \bibinfo{person}{Eitan Grinspun}.} \bibinfo{year}{2010}\natexlab{}.
\newblock \showarticletitle{Discrete viscous threads}. In
  \bibinfo{booktitle}{\emph{ACM TOG}}.
\newblock


\bibitem[\protect\citeauthoryear{Bergou, Wardetzky, Robinson, Audoly, and
  Grinspun}{Bergou et~al\mbox{.}}{2008}]%
        {bergou2008discrete}
\bibfield{author}{\bibinfo{person}{Mikl{\'o}s Bergou}, \bibinfo{person}{Max
  Wardetzky}, \bibinfo{person}{Stephen Robinson}, \bibinfo{person}{Basile
  Audoly}, {and} \bibinfo{person}{Eitan Grinspun}.}
  \bibinfo{year}{2008}\natexlab{}.
\newblock \showarticletitle{Discrete elastic rods}. In
  \bibinfo{booktitle}{\emph{ACM TOG}}.
\newblock


\bibitem[\protect\citeauthoryear{Bertails, Audoly, Cani, Querleux, Leroy, and
  L{\'e}v{\^e}que}{Bertails et~al\mbox{.}}{2006}]%
        {bertails2006super}
\bibfield{author}{\bibinfo{person}{Florence Bertails}, \bibinfo{person}{Basile
  Audoly}, \bibinfo{person}{Marie-Paule Cani}, \bibinfo{person}{Bernard
  Querleux}, \bibinfo{person}{Fr{\'e}d{\'e}ric Leroy}, {and}
  \bibinfo{person}{Jean-Luc L{\'e}v{\^e}que}.} \bibinfo{year}{2006}\natexlab{}.
\newblock \showarticletitle{Super-helices for predicting the dynamics of
  natural hair}. In \bibinfo{booktitle}{\emph{ACM TOG}}.
\newblock


\bibitem[\protect\citeauthoryear{Botsch, Kobbelt, Pauly, Alliez, and
  L{\'e}vy}{Botsch et~al\mbox{.}}{2010}]%
        {pmp}
\bibfield{author}{\bibinfo{person}{Mario Botsch}, \bibinfo{person}{Leif
  Kobbelt}, \bibinfo{person}{Mark Pauly}, \bibinfo{person}{Pierre Alliez},
  {and} \bibinfo{person}{Bruno L{\'e}vy}.} \bibinfo{year}{2010}\natexlab{}.
\newblock \bibinfo{booktitle}{\emph{Polygon mesh processing}}.
\newblock \bibinfo{publisher}{AK Peters/CRC Press}.
\newblock


\bibitem[\protect\citeauthoryear{Bouaziz, Martin, Liu, Kavan, and
  Pauly}{Bouaziz et~al\mbox{.}}{2014}]%
        {projdyn}
\bibfield{author}{\bibinfo{person}{Sofien Bouaziz}, \bibinfo{person}{Sebastian
  Martin}, \bibinfo{person}{Tiantian Liu}, \bibinfo{person}{Ladislav Kavan},
  {and} \bibinfo{person}{Mark Pauly}.} \bibinfo{year}{2014}\natexlab{}.
\newblock \showarticletitle{Projective Dynamics: Fusing Constraint Projections
  for Fast Simulation}.
\newblock \bibinfo{journal}{\emph{ACM TOG}} (\bibinfo{year}{2014}).
\newblock


\bibitem[\protect\citeauthoryear{Choi and Blemker}{Choi and Blemker}{2013}]%
        {choi_plos13}
\bibfield{author}{\bibinfo{person}{Hon~Fai Choi} {and}
  \bibinfo{person}{Silvia~S Blemker}.} \bibinfo{year}{2013}\natexlab{}.
\newblock \showarticletitle{Skeletal muscle fascicle arrangements can be
  reconstructed using a laplacian vector field simulation}.
\newblock \bibinfo{journal}{\emph{PloS one}} (\bibinfo{year}{2013}).
\newblock


\bibitem[\protect\citeauthoryear{Clutterbuck and Jacobs}{Clutterbuck and
  Jacobs}{2010}]%
        {avatar}
\bibfield{author}{\bibinfo{person}{Simon Clutterbuck} {and}
  \bibinfo{person}{James Jacobs}.} \bibinfo{year}{2010}\natexlab{}.
\newblock \bibinfo{title}{A Physically Based Approach to Virtual Character
  Deformations}.
\newblock \bibinfo{howpublished}{In ACM SIGGRAPH Talk sessions}.
\newblock


\bibitem[\protect\citeauthoryear{Comet}{Comet}{2011}]%
        {mayamuscle}
\bibfield{author}{\bibinfo{person}{Michael Comet}.}
  \bibinfo{year}{2011}\natexlab{}.
\newblock \bibinfo{title}{Maya Muscle}.
\newblock
  \bibinfo{howpublished}{\url{http://download.autodesk.com/us/support/files/muscle.pdf}}.
\newblock
\newblock
\shownote{(Accessed on Aug. 8th, 2018).}


\bibitem[\protect\citeauthoryear{Gould}{Gould}{1986}]%
        {gould1986accurate}
\bibfield{author}{\bibinfo{person}{Nicholas Ian~Mark Gould}.}
  \bibinfo{year}{1986}\natexlab{}.
\newblock \showarticletitle{On the accurate determination of search directions
  for simple differentiable penalty functions}.
\newblock \bibinfo{journal}{\emph{IMA J. Numer. Anal.}} (\bibinfo{year}{1986}).
\newblock


\bibitem[\protect\citeauthoryear{Green}{Green}{2010}]%
        {green2010particle}
\bibfield{author}{\bibinfo{person}{Simon Green}.}
  \bibinfo{year}{2010}\natexlab{}.
\newblock \showarticletitle{Particle simulation using {CUDA}}.
\newblock \bibinfo{journal}{\emph{NVIDIA whitepaper}} (\bibinfo{year}{2010}).
\newblock


\bibitem[\protect\citeauthoryear{Gr{\'e}goire and Sch\"{o}mer}{Gr{\'e}goire and
  Sch\"{o}mer}{2006}]%
        {flexible_2006}
\bibfield{author}{\bibinfo{person}{Mireille Gr{\'e}goire} {and}
  \bibinfo{person}{Elmar Sch\"{o}mer}.} \bibinfo{year}{2006}\natexlab{}.
\newblock \showarticletitle{Interactive Simulation of One-dimensional Flexible
  Parts}. In \bibinfo{booktitle}{\emph{Proc. of ACM Symposium on Solid and
  Physical Modeling}}.
\newblock


\bibitem[\protect\citeauthoryear{Ichim, Kadle{\v{c}}ek, Kavan, and Pauly}{Ichim
  et~al\mbox{.}}{2017}]%
        {ichim2017phace}
\bibfield{author}{\bibinfo{person}{Alexandru-Eugen Ichim},
  \bibinfo{person}{Petr Kadle{\v{c}}ek}, \bibinfo{person}{Ladislav Kavan},
  {and} \bibinfo{person}{Mark Pauly}.} \bibinfo{year}{2017}\natexlab{}.
\newblock \showarticletitle{Phace: Physics-based face modeling and animation}.
\newblock \bibinfo{journal}{\emph{ACM TOG}} (\bibinfo{year}{2017}).
\newblock


\bibitem[\protect\citeauthoryear{Jacobson, Deng, Kavan, and Lewis}{Jacobson
  et~al\mbox{.}}{2014}]%
        {skinning}
\bibfield{author}{\bibinfo{person}{Alec Jacobson}, \bibinfo{person}{Zhigang
  Deng}, \bibinfo{person}{Ladislav Kavan}, {and} \bibinfo{person}{J.P. Lewis}.}
  \bibinfo{year}{2014}\natexlab{}.
\newblock \bibinfo{title}{Skinning: Real-time Shape Deformation}.
\newblock \bibinfo{howpublished}{SIGGRAPH Course,
  http://skinning.org/direct-methods.pdf}.
\newblock


\bibitem[\protect\citeauthoryear{Kadle{\v{c}}ek, Ichim, Liu,
  K{\v{r}}iv{\'a}nek, and Kavan}{Kadle{\v{c}}ek et~al\mbox{.}}{2016}]%
        {kadlecek_siga16}
\bibfield{author}{\bibinfo{person}{Petr Kadle{\v{c}}ek},
  \bibinfo{person}{Alexandru-Eugen Ichim}, \bibinfo{person}{Tiantian Liu},
  \bibinfo{person}{Jaroslav K{\v{r}}iv{\'a}nek}, {and}
  \bibinfo{person}{Ladislav Kavan}.} \bibinfo{year}{2016}\natexlab{}.
\newblock \showarticletitle{Reconstructing personalized anatomical models for
  physics-based body animation}.
\newblock \bibinfo{journal}{\emph{ACM TOG (Proc. SIGGRAPH Asia)}}
  (\bibinfo{year}{2016}).
\newblock


\bibitem[\protect\citeauthoryear{Kavan, Collins, \v{Z}\'{a}ra, and
  O'Sullivan}{Kavan et~al\mbox{.}}{2007}]%
        {dqs}
\bibfield{author}{\bibinfo{person}{Ladislav Kavan}, \bibinfo{person}{Steven
  Collins}, \bibinfo{person}{Ji\v{r}\'{\i} \v{Z}\'{a}ra}, {and}
  \bibinfo{person}{Carol O'Sullivan}.} \bibinfo{year}{2007}\natexlab{}.
\newblock \showarticletitle{Skinning with Dual Quaternions}. In
  \bibinfo{booktitle}{\emph{Proceedings of the 2007 Symposium on Interactive 3D
  Graphics and Games}}.
\newblock


\bibitem[\protect\citeauthoryear{Kugelstadt and Sch{\"o}mer}{Kugelstadt and
  Sch{\"o}mer}{2016}]%
        {pbdcorods}
\bibfield{author}{\bibinfo{person}{Tassilo Kugelstadt} {and}
  \bibinfo{person}{Elmar Sch{\"o}mer}.} \bibinfo{year}{2016}\natexlab{}.
\newblock \showarticletitle{Position and orientation based Cosserat rods.}. In
  \bibinfo{booktitle}{\emph{Proc. SCA}}.
\newblock


\bibitem[\protect\citeauthoryear{Kugelstadt and Sch\"{o}mer}{Kugelstadt and
  Sch\"{o}mer}{2016}]%
        {Kugelstadt:2016}
\bibfield{author}{\bibinfo{person}{T. Kugelstadt} {and} \bibinfo{person}{E.
  Sch\"{o}mer}.} \bibinfo{year}{2016}\natexlab{}.
\newblock \showarticletitle{Position and Orientation Based Cosserat Rods}. In
  \bibinfo{booktitle}{\emph{Proc. SCA}}.
\newblock


\bibitem[\protect\citeauthoryear{Kurihara and Miyata}{Kurihara and
  Miyata}{2004}]%
        {kurihara04}
\bibfield{author}{\bibinfo{person}{Tsuneya Kurihara} {and}
  \bibinfo{person}{Natsuki Miyata}.} \bibinfo{year}{2004}\natexlab{}.
\newblock \showarticletitle{Modeling Deformable Human Hands from Medical
  Images}. In \bibinfo{booktitle}{\emph{Proceedings of the 2004 {ACM}
  {SIGGRAPH} Symposium on Computer Animation ({SCA}-04)}}.
\newblock


\bibitem[\protect\citeauthoryear{Lang, Linn, and Arnold}{Lang
  et~al\mbox{.}}{2011}]%
        {exact_2011}
\bibfield{author}{\bibinfo{person}{Holger Lang}, \bibinfo{person}{Joachim
  Linn}, {and} \bibinfo{person}{Martin Arnold}.}
  \bibinfo{year}{2011}\natexlab{}.
\newblock \showarticletitle{Multi-body dynamics simulation of geometrically
  exact Cosserat rods}.
\newblock \bibinfo{journal}{\emph{Multibody System Dynamics}}
  (\bibinfo{year}{2011}).
\newblock


\bibitem[\protect\citeauthoryear{Le and Hodgins}{Le and Hodgins}{2016}]%
        {binhleCoR}
\bibfield{author}{\bibinfo{person}{Binh~Huy Le} {and}
  \bibinfo{person}{Jessica~K. Hodgins}.} \bibinfo{year}{2016}\natexlab{}.
\newblock \showarticletitle{Real-time Skeletal Skinning with Optimized Centers
  of Rotation}.
\newblock \bibinfo{journal}{\emph{ACM Trans. Graph.}} (\bibinfo{year}{2016}).
\newblock


\bibitem[\protect\citeauthoryear{Lee, Glueck, Khan, Fiume, and Jackson}{Lee
  et~al\mbox{.}}{2010}]%
        {lee_tog10}
\bibfield{author}{\bibinfo{person}{Dongwoon Lee}, \bibinfo{person}{Michael
  Glueck}, \bibinfo{person}{Azam Khan}, \bibinfo{person}{Eugene Fiume}, {and}
  \bibinfo{person}{Ken Jackson}.} \bibinfo{year}{2010}\natexlab{}.
\newblock \showarticletitle{A survey of modeling and simulation of skeletal
  muscle}.
\newblock \bibinfo{journal}{\emph{ACM TOG}} (\bibinfo{year}{2010}).
\newblock


\bibitem[\protect\citeauthoryear{Lewis, Anjyo, Rhee, Zhang, Pighin, and
  Deng}{Lewis et~al\mbox{.}}{2014}]%
        {blendshapes}
\bibfield{author}{\bibinfo{person}{J.P. Lewis}, \bibinfo{person}{Ken Anjyo},
  \bibinfo{person}{Taehyun Rhee}, \bibinfo{person}{Mengjie Zhang},
  \bibinfo{person}{Fred Pighin}, {and} \bibinfo{person}{Zhigang Deng}.}
  \bibinfo{year}{2014}\natexlab{}.
\newblock \showarticletitle{STAR: Practice and Theory of Blendshape Facial
  Models}. In \bibinfo{booktitle}{\emph{Eurographics}}.
\newblock


\bibitem[\protect\citeauthoryear{Lewis, Cordner, and Fong}{Lewis
  et~al\mbox{.}}{2000}]%
        {psd}
\bibfield{author}{\bibinfo{person}{J.~P. Lewis}, \bibinfo{person}{Matt
  Cordner}, {and} \bibinfo{person}{Nickson Fong}.}
  \bibinfo{year}{2000}\natexlab{}.
\newblock \showarticletitle{Pose Space Deformation: A Unified Approach to Shape
  Interpolation and Skeleton-Driven Deformation}. In
  \bibinfo{booktitle}{\emph{Proc. ACM SIGGRAPH}}.
\newblock


\bibitem[\protect\citeauthoryear{Li, Sueda, Neog, and Pai}{Li
  et~al\mbox{.}}{2013}]%
        {thinskin}
\bibfield{author}{\bibinfo{person}{Duo Li}, \bibinfo{person}{Shinjiro Sueda},
  \bibinfo{person}{Debanga~R Neog}, {and} \bibinfo{person}{Dinesh~K Pai}.}
  \bibinfo{year}{2013}\natexlab{}.
\newblock \showarticletitle{Thin Skin Elastodynamics}.
\newblock \bibinfo{journal}{\emph{ACM TOG (Proc. SIGGRAPH)}}
  (\bibinfo{year}{2013}).
\newblock


\bibitem[\protect\citeauthoryear{Loper, Mahmood, Romero, Pons-Moll, and
  Black}{Loper et~al\mbox{.}}{2015}]%
        {smpl}
\bibfield{author}{\bibinfo{person}{Matthew Loper}, \bibinfo{person}{Naureen
  Mahmood}, \bibinfo{person}{Javier Romero}, \bibinfo{person}{Gerard
  Pons-Moll}, {and} \bibinfo{person}{Michael~J Black}.}
  \bibinfo{year}{2015}\natexlab{}.
\newblock \showarticletitle{SMPL: A skinned multi-person linear model}.
\newblock \bibinfo{journal}{\emph{ACM TOG}} (\bibinfo{year}{2015}).
\newblock


\bibitem[\protect\citeauthoryear{Macklin, M\"{u}ller, and Chentanez}{Macklin
  et~al\mbox{.}}{2016}]%
        {xpbd}
\bibfield{author}{\bibinfo{person}{Miles Macklin}, \bibinfo{person}{Matthias
  M\"{u}ller}, {and} \bibinfo{person}{Nuttapong Chentanez}.}
  \bibinfo{year}{2016}\natexlab{}.
\newblock \showarticletitle{{XPBD: Position-based Simulation of Compliant
  Constrained Dynamics}}. In \bibinfo{booktitle}{\emph{Proc. of the
  International Conference on Motion in Games}}.
\newblock


\bibitem[\protect\citeauthoryear{Macklin, M{\"u}ller, Chentanez, and
  Kim}{Macklin et~al\mbox{.}}{2014}]%
        {flex}
\bibfield{author}{\bibinfo{person}{Miles Macklin}, \bibinfo{person}{Matthias
  M{\"u}ller}, \bibinfo{person}{Nuttapong Chentanez}, {and}
  \bibinfo{person}{Tae-Yong Kim}.} \bibinfo{year}{2014}\natexlab{}.
\newblock \showarticletitle{Unified particle physics for real-time
  applications}.
\newblock \bibinfo{journal}{\emph{ACM TOG (Proc. SIGGRAPH)}}
  (\bibinfo{year}{2014}).
\newblock


\bibitem[\protect\citeauthoryear{Martin, Thomaszewski, Grinspun, and
  Gross}{Martin et~al\mbox{.}}{2011}]%
        {martin_tog11}
\bibfield{author}{\bibinfo{person}{Sebastian Martin}, \bibinfo{person}{Bernhard
  Thomaszewski}, \bibinfo{person}{Eitan Grinspun}, {and}
  \bibinfo{person}{Markus Gross}.} \bibinfo{year}{2011}\natexlab{}.
\newblock \showarticletitle{Example-based Elastic Materials}.
\newblock \bibinfo{journal}{\emph{ACM TOG}} (\bibinfo{year}{2011}).
\newblock


\bibitem[\protect\citeauthoryear{Muller}{Muller}{2008}]%
        {physx}
\bibfield{author}{\bibinfo{person}{Matthias Muller}.}
  \bibinfo{year}{2008}\natexlab{}.
\newblock \bibinfo{title}{{NVIDIA PhysX SDK} 3.4.0 Documentation}.
\newblock
  \bibinfo{howpublished}{\url{https://docs.nvidia.com/gameworks/\#gameworkslibrary/physx/physx.htm}}.
\newblock
\newblock
\shownote{(Accessed on Aug. 9th, 2018).}


\bibitem[\protect\citeauthoryear{M\"{u}ller, Bender, Chentanez, and
  Macklin}{M\"{u}ller et~al\mbox{.}}{2016}]%
        {Muller:2016}
\bibfield{author}{\bibinfo{person}{Matthias M\"{u}ller}, \bibinfo{person}{Jan
  Bender}, \bibinfo{person}{Nuttapong Chentanez}, {and} \bibinfo{person}{Miles
  Macklin}.} \bibinfo{year}{2016}\natexlab{}.
\newblock \showarticletitle{A Robust Method to Extract the Rotational Part of
  Deformations}. In \bibinfo{booktitle}{\emph{Proceedings of the 9th
  International Conference on Motion in Games}} \emph{(\bibinfo{series}{MIG
  '16})}.
\newblock


\bibitem[\protect\citeauthoryear{M{\"u}ller and Chentanez}{M{\"u}ller and
  Chentanez}{2011}]%
        {muller_vriphys11}
\bibfield{author}{\bibinfo{person}{Matthias M{\"u}ller} {and}
  \bibinfo{person}{Nuttapong Chentanez}.} \bibinfo{year}{2011}\natexlab{}.
\newblock \showarticletitle{Adding Physics to Animated Characters with Oriented
  Particles}.
\newblock


\bibitem[\protect\citeauthoryear{M\"{u}ller and Chentanez}{M\"{u}ller and
  Chentanez}{2011}]%
        {muller_tog11}
\bibfield{author}{\bibinfo{person}{Matthias M\"{u}ller} {and}
  \bibinfo{person}{Nuttapong Chentanez}.} \bibinfo{year}{2011}\natexlab{}.
\newblock \showarticletitle{Solid simulation with oriented particles}.
\newblock \bibinfo{journal}{\emph{ACM TOG}} (\bibinfo{year}{2011}).
\newblock


\bibitem[\protect\citeauthoryear{M{\"u}ller, Heidelberger, Hennix, and
  Ratcliff}{M{\"u}ller et~al\mbox{.}}{2007}]%
        {pbd}
\bibfield{author}{\bibinfo{person}{Matthias M{\"u}ller}, \bibinfo{person}{Bruno
  Heidelberger}, \bibinfo{person}{Marcus Hennix}, {and} \bibinfo{person}{John
  Ratcliff}.} \bibinfo{year}{2007}\natexlab{}.
\newblock \showarticletitle{Position based dynamics}.
\newblock \bibinfo{journal}{\emph{Journal of Visual Communication and Image
  Representation}} (\bibinfo{year}{2007}).
\newblock


\bibitem[\protect\citeauthoryear{Pai}{Pai}{2002}]%
        {strands}
\bibfield{author}{\bibinfo{person}{Dinesh~K Pai}.}
  \bibinfo{year}{2002}\natexlab{}.
\newblock \showarticletitle{Strands: Interactive simulation of thin solids
  using cosserat models}. In \bibinfo{booktitle}{\emph{Computer Graphics
  Forum}}.
\newblock


\bibitem[\protect\citeauthoryear{Pons-Moll, Romero, Mahmood, and
  Black}{Pons-Moll et~al\mbox{.}}{2015}]%
        {dyna}
\bibfield{author}{\bibinfo{person}{Gerard Pons-Moll}, \bibinfo{person}{Javier
  Romero}, \bibinfo{person}{Naureen Mahmood}, {and} \bibinfo{person}{Michael~J.
  Black}.} \bibinfo{year}{2015}\natexlab{}.
\newblock \showarticletitle{Dyna: A Model of Dynamic Human Shape in Motion}.
\newblock \bibinfo{journal}{\emph{ACM TOG (Proc. SIGGRAPH)}}
  (\bibinfo{year}{2015}).
\newblock


\bibitem[\protect\citeauthoryear{Romeo, Monteagudo, and
  S\'anchez-Quir\'os}{Romeo et~al\mbox{.}}{2018}]%
        {romeo_ceig18}
\bibfield{author}{\bibinfo{person}{Marco Romeo}, \bibinfo{person}{Carlos
  Monteagudo}, {and} \bibinfo{person}{Daniel S\'anchez-Quir\'os}.}
  \bibinfo{year}{2018}\natexlab{}.
\newblock \showarticletitle{{Muscle Simulation with Extended Position Based
  Dynamics}}. In \bibinfo{booktitle}{\emph{Spanish Computer Graphics Conference
  (CEIG)}}.
\newblock


\bibitem[\protect\citeauthoryear{Saito and Yuen}{Saito and Yuen}{2017}]%
        {saitoskinslide}
\bibfield{author}{\bibinfo{person}{Jun Saito} {and} \bibinfo{person}{Simon
  Yuen}.} \bibinfo{year}{2017}\natexlab{}.
\newblock \showarticletitle{Efficient and Robust Skin Slide Simulation}. In
  \bibinfo{booktitle}{\emph{Proceedings of the ACM SIGGRAPH Digital Production
  Symposium}} \emph{(\bibinfo{series}{DigiPro '17})}.
\newblock


\bibitem[\protect\citeauthoryear{Saito, Zhou, and Kavan}{Saito
  et~al\mbox{.}}{2015}]%
        {saito_sig15}
\bibfield{author}{\bibinfo{person}{Shunsuke Saito}, \bibinfo{person}{Zi-Ye
  Zhou}, {and} \bibinfo{person}{Ladislav Kavan}.}
  \bibinfo{year}{2015}\natexlab{}.
\newblock \showarticletitle{Computational bodybuilding: Anatomically-based
  modeling of human bodies}.
\newblock \bibinfo{journal}{\emph{ACM TOG (Proc. SIGGRAPH)}}
  (\bibinfo{year}{2015}).
\newblock


\bibitem[\protect\citeauthoryear{Scheepers, Parent, Carlson, and May}{Scheepers
  et~al\mbox{.}}{1997}]%
        {scheepers97}
\bibfield{author}{\bibinfo{person}{Ferdi Scheepers},
  \bibinfo{person}{Richard~E. Parent}, \bibinfo{person}{Wayne~E. Carlson},
  {and} \bibinfo{person}{Stephen~F. May}.} \bibinfo{year}{1997}\natexlab{}.
\newblock \showarticletitle{Anatomy-Based Modeling of the Human Musculature}
  \emph{(\bibinfo{series}{SIGGRAPH '97})}.
\newblock


\bibitem[\protect\citeauthoryear{Schumacher, Thomaszewski, Coros, Martin,
  Sumner, and Gross}{Schumacher et~al\mbox{.}}{2012}]%
        {examplemat}
\bibfield{author}{\bibinfo{person}{Christian Schumacher},
  \bibinfo{person}{Bernhard Thomaszewski}, \bibinfo{person}{Stelian Coros},
  \bibinfo{person}{Sebastian Martin}, \bibinfo{person}{Robert Sumner}, {and}
  \bibinfo{person}{Markus Gross}.} \bibinfo{year}{2012}\natexlab{}.
\newblock \showarticletitle{Efficient simulation of example-based materials}.
  In \bibinfo{booktitle}{\emph{Proceedings of the 11th ACM
  SIGGRAPH/Eurographics conference on Computer Animation}}.
\newblock


\bibitem[\protect\citeauthoryear{Sifakis and Barbic}{Sifakis and
  Barbic}{2012}]%
        {Sifakis:2012}
\bibfield{author}{\bibinfo{person}{Eftychios Sifakis} {and}
  \bibinfo{person}{Jernej Barbic}.} \bibinfo{year}{2012}\natexlab{}.
\newblock \showarticletitle{FEM Simulation of 3D Deformable Solids: A
  Practitioner's Guide to Theory, Discretization and Model Reduction}. In
  \bibinfo{booktitle}{\emph{ACM SIGGRAPH 2012 Courses}}
  \emph{(\bibinfo{series}{SIGGRAPH '12})}.
\newblock


\bibitem[\protect\citeauthoryear{Sifakis, Neverov, and Fedkiw}{Sifakis
  et~al\mbox{.}}{2005}]%
        {sifakis05}
\bibfield{author}{\bibinfo{person}{Eftychios Sifakis}, \bibinfo{person}{Igor
  Neverov}, {and} \bibinfo{person}{Ronald Fedkiw}.}
  \bibinfo{year}{2005}\natexlab{}.
\newblock \showarticletitle{Automatic Determination of Facial Muscle
  Activations from Sparse Motion Capture Marker Data}. In
  \bibinfo{booktitle}{\emph{ACM TOG (Proc. SIGGRAPH)}}.
\newblock


\bibitem[\protect\citeauthoryear{Smith, Goes, and Kim}{Smith
  et~al\mbox{.}}{2018}]%
        {smith2018stable}
\bibfield{author}{\bibinfo{person}{Breannan Smith},
  \bibinfo{person}{Fernando~De Goes}, {and} \bibinfo{person}{Theodore Kim}.}
  \bibinfo{year}{2018}\natexlab{}.
\newblock \showarticletitle{Stable Neo-Hookean Flesh Simulation}.
\newblock \bibinfo{journal}{\emph{ACM TOG}} \bibinfo{volume}{37},
  \bibinfo{number}{2} (\bibinfo{year}{2018}), \bibinfo{pages}{12}.
\newblock


\bibitem[\protect\citeauthoryear{Soler, Martin, and Sorkine-Hornung}{Soler
  et~al\mbox{.}}{2018}]%
        {soler2018cosserat}
\bibfield{author}{\bibinfo{person}{Carlota Soler}, \bibinfo{person}{Tobias
  Martin}, {and} \bibinfo{person}{Olga Sorkine-Hornung}.}
  \bibinfo{year}{2018}\natexlab{}.
\newblock \showarticletitle{Cosserat Rods with Projective Dynamics}. In
  \bibinfo{booktitle}{\emph{Computer Graphics Forum}}.
\newblock


\bibitem[\protect\citeauthoryear{Spillmann and Teschner}{Spillmann and
  Teschner}{2007}]%
        {rods}
\bibfield{author}{\bibinfo{person}{J. Spillmann} {and} \bibinfo{person}{M.
  Teschner}.} \bibinfo{year}{2007}\natexlab{}.
\newblock \showarticletitle{CORDE: Cosserat Rod Elements for the Dynamic
  Simulation of One-Dimensional Elastic Objects}. In
  \bibinfo{booktitle}{\emph{Proc. SCA}}.
\newblock


\bibitem[\protect\citeauthoryear{Sueda, Jones, Levin, and Pai}{Sueda
  et~al\mbox{.}}{2011}]%
        {sueda2011large}
\bibfield{author}{\bibinfo{person}{Shinjiro Sueda}, \bibinfo{person}{Garrett~L
  Jones}, \bibinfo{person}{David~IW Levin}, {and} \bibinfo{person}{Dinesh~K
  Pai}.} \bibinfo{year}{2011}\natexlab{}.
\newblock \showarticletitle{Large-scale dynamic simulation of highly
  constrained strands}.
\newblock \bibinfo{journal}{\emph{ACM TOG}} (\bibinfo{year}{2011}).
\newblock


\bibitem[\protect\citeauthoryear{Sueda, Kaufman, and Pai}{Sueda
  et~al\mbox{.}}{2008}]%
        {sueda2008musculotendon}
\bibfield{author}{\bibinfo{person}{Shinjiro Sueda}, \bibinfo{person}{Andrew
  Kaufman}, {and} \bibinfo{person}{Dinesh~K Pai}.}
  \bibinfo{year}{2008}\natexlab{}.
\newblock \showarticletitle{Musculotendon simulation for hand animation}.
\newblock \bibinfo{journal}{\emph{ACM TOG}} (\bibinfo{year}{2008}).
\newblock


\bibitem[\protect\citeauthoryear{Tagliasacchi, Delame, Spagnuolo, Amenta, and
  Telea}{Tagliasacchi et~al\mbox{.}}{2016}]%
        {skelstar_eg16}
\bibfield{author}{\bibinfo{person}{Andrea Tagliasacchi},
  \bibinfo{person}{Thomas Delame}, \bibinfo{person}{Michela Spagnuolo},
  \bibinfo{person}{Nina Amenta}, {and} \bibinfo{person}{Alexandru Telea}.}
  \bibinfo{year}{2016}\natexlab{}.
\newblock \showarticletitle{3D Skeletons: A State-of-the-Art Report}.
\newblock \bibinfo{journal}{\emph{Proc. Eurographics (State of the Art
  Reports)}} (\bibinfo{year}{2016}).
\newblock


\bibitem[\protect\citeauthoryear{Terzopoulos and Waters}{Terzopoulos and
  Waters}{1990}]%
        {terzopoulos90}
\bibfield{author}{\bibinfo{person}{Demetri Terzopoulos} {and}
  \bibinfo{person}{Keith Waters}.} \bibinfo{year}{1990}\natexlab{}.
\newblock \showarticletitle{Physically-based Facial Modeling, Analysis, and
  Animation}.
\newblock \bibinfo{journal}{\emph{Journal of Visualization and Computer
  Animation}} (\bibinfo{year}{1990}).
\newblock


\bibitem[\protect\citeauthoryear{Thiery, Guy, and Boubekeur}{Thiery
  et~al\mbox{.}}{2013}]%
        {sphmsh}
\bibfield{author}{\bibinfo{person}{Jean-Marc Thiery},
  \bibinfo{person}{{\'E}milie Guy}, {and} \bibinfo{person}{Tamy Boubekeur}.}
  \bibinfo{year}{2013}\natexlab{}.
\newblock \showarticletitle{Sphere-Meshes: shape approximation using spherical
  quadric error metrics}.
\newblock \bibinfo{journal}{\emph{ACM TOG}} (\bibinfo{year}{2013}).
\newblock


\bibitem[\protect\citeauthoryear{Tkach, Pauly, and Tagliasacchi}{Tkach
  et~al\mbox{.}}{2016}]%
        {hmodel}
\bibfield{author}{\bibinfo{person}{Anastasia Tkach}, \bibinfo{person}{Mark
  Pauly}, {and} \bibinfo{person}{Andrea Tagliasacchi}.}
  \bibinfo{year}{2016}\natexlab{}.
\newblock \showarticletitle{Sphere-Meshes for Real-Time Hand Modeling and
  Tracking}.
\newblock \bibinfo{journal}{\emph{ACM TOG (Proc. SIGGRAPH Asia)}}
  (\bibinfo{year}{2016}).
\newblock


\bibitem[\protect\citeauthoryear{Tkach, Tagliasacchi, Remelli, Pauly, and
  Fitzgibbon}{Tkach et~al\mbox{.}}{2017}]%
        {honline}
\bibfield{author}{\bibinfo{person}{Anastasia Tkach}, \bibinfo{person}{Andrea
  Tagliasacchi}, \bibinfo{person}{Edoardo Remelli}, \bibinfo{person}{Mark
  Pauly}, {and} \bibinfo{person}{Andrew Fitzgibbon}.}
  \bibinfo{year}{2017}\natexlab{}.
\newblock \showarticletitle{Online Generative Model Personalization for Hand
  Tracking}.
\newblock \bibinfo{journal}{\emph{ACM TOG (Proc. SIGGRAPH Asia)}}
  (\bibinfo{year}{2017}).
\newblock


\bibitem[\protect\citeauthoryear{Umetani, Schmidt, and Stam}{Umetani
  et~al\mbox{.}}{2014}]%
        {pbdrods}
\bibfield{author}{\bibinfo{person}{Nobuyuki Umetani}, \bibinfo{person}{Ryan
  Schmidt}, {and} \bibinfo{person}{Jos Stam}.} \bibinfo{year}{2014}\natexlab{}.
\newblock \showarticletitle{Position-based elastic rods}. In
  \bibinfo{booktitle}{\emph{Proc. SCA}}.
\newblock


\bibitem[\protect\citeauthoryear{Umeyama}{Umeyama}{1991}]%
        {umeyama:1991}
\bibfield{author}{\bibinfo{person}{Shinji Umeyama}.}
  \bibinfo{year}{1991}\natexlab{}.
\newblock \showarticletitle{Least-Squares Estimation of Transformation
  Parameters Between Two Point Patterns}.
\newblock \bibinfo{journal}{\emph{IEEE Trans. Pattern Anal. Mach. Intell.}}
  (\bibinfo{year}{1991}).
\newblock


\bibitem[\protect\citeauthoryear{Vaillant, Barthe, Guennebaud, Cani, Rohmer,
  Wyvill, Gourmel, and Paulin}{Vaillant et~al\mbox{.}}{2013}]%
        {impski}
\bibfield{author}{\bibinfo{person}{Rodolphe Vaillant},
  \bibinfo{person}{Lo\"{\i}c Barthe}, \bibinfo{person}{Ga\"{e}l Guennebaud},
  \bibinfo{person}{Marie-Paule Cani}, \bibinfo{person}{Damien Rohmer},
  \bibinfo{person}{Brian Wyvill}, \bibinfo{person}{Olivier Gourmel}, {and}
  \bibinfo{person}{Mathias Paulin}.} \bibinfo{year}{2013}\natexlab{}.
\newblock \showarticletitle{Implicit Skinning: Real-time Skin Deformation with
  Contact Modeling}.
\newblock \bibinfo{journal}{\emph{ACM TOG (Proc. SIGGRAPH)}}
  (\bibinfo{year}{2013}).
\newblock


\bibitem[\protect\citeauthoryear{Vital Mechanics}{Vital Mechanics}{2018}]%
        {vital}
Vital Mechanics \bibinfo{year}{2018}\natexlab{}.
\newblock \bibinfo{howpublished}{\textsf{http://www.vital.com}}.
\newblock


\bibitem[\protect\citeauthoryear{Xu and Barbi\v{c}}{Xu and Barbi\v{c}}{2016}]%
        {pssd}
\bibfield{author}{\bibinfo{person}{Hongyi Xu} {and} \bibinfo{person}{Jernej
  Barbi\v{c}}.} \bibinfo{year}{2016}\natexlab{}.
\newblock \showarticletitle{Pose-space Subspace Dynamics}.
\newblock \bibinfo{journal}{\emph{ACM TOG}} (\bibinfo{year}{2016}).
\newblock


\bibitem[\protect\citeauthoryear{Yuen}{Yuen}{2018}]%
        {artistquote}
\bibfield{author}{\bibinfo{person}{Simon Yuen}.}
  \bibinfo{year}{2018}\natexlab{}.
\newblock \bibinfo{title}{\emph{Personal communication}}.
\newblock \bibinfo{howpublished}{Head of Creatures, Method Studios}.
\newblock


\bibitem[\protect\citeauthoryear{Zhu, Hu, and Kavan}{Zhu et~al\mbox{.}}{2015}]%
        {zhu_eg15}
\bibfield{author}{\bibinfo{person}{Lifeng Zhu}, \bibinfo{person}{Xiaoyan Hu},
  {and} \bibinfo{person}{Ladislav Kavan}.} \bibinfo{year}{2015}\natexlab{}.
\newblock \showarticletitle{Adaptable anatomical models for realistic bone
  motion reconstruction}.
\newblock \bibinfo{journal}{\emph{CGF (Proc. EuroGraphics)}}
  (\bibinfo{year}{2015}).
\newblock


\bibitem[\protect\citeauthoryear{Ziva Dynamics}{Ziva Dynamics}{2018}]%
        {ziva}
Ziva Dynamics \bibinfo{year}{2018}\natexlab{}.
\newblock \bibinfo{title}{Ziva Dynamics}.
\newblock \bibinfo{howpublished}{\textsf{https://zivadynamics.com}}.
\newblock


\end{thebibliography}
